\documentclass[nofootinbib,aps,a4paper,letterpaper,superscriptaddress,
twocolumn,showpacs,times,eqsecnum]{revtex4}
\usepackage{amsmath}
\usepackage{amsfonts}
\usepackage{booktabs}
\usepackage{multirow}
\usepackage{siunitx} 
\usepackage{adjustbox}
\pdfoutput=1
\usepackage{graphicx}
\usepackage{color}
\usepackage{braket}
\usepackage{dcolumn}
\usepackage{bm,url}
\usepackage[linktocpage]{hyperref}
\usepackage{subfigure}
\usepackage{amsfonts}
\usepackage[usenames,dvipsnames,svgnames]{xcolor}  %% colored text
\usepackage{hyperref}   
\definecolor{oxfordblue}{rgb}{0.0, 0.13, 0.28}
\definecolor{burgundy}{rgb}{0.5, 0.0, 0.13}
\definecolor{darkolivegreen}{rgb}{0.33, 0.42, 0.18}
\definecolor{darkblue}{rgb}{0,0,0.5}
\definecolor{richcarmine}{rgb}{0.84, 0.0, 0.25}
\definecolor{darkblue}{rgb}{0,0,0.5}
\definecolor{bluer}{rgb}{0.00,0.50,0.75}{}
\hypersetup{colorlinks=true, citecolor=red, linkcolor=blue,
 urlcolor = magenta, filecolor=magenta}

\begin{document}
 
 \newcommand\be{\begin{equation}}
  \newcommand\ee{\end{equation}}
 \newcommand\bea{\begin{eqnarray}}
  \newcommand\eea{\end{eqnarray}}
 \newcommand\bseq{\begin{subequations}} %solo con amsmath
  \newcommand\eseq{\end{subequations}}
 \newcommand\bcas{\begin{cases}}
  \newcommand\ecas{\end{cases}}
 \newcommand{\p}{\partial}
 \newcommand{\f}{\frac}

 \title{Cosmological evolution and late-time attractors in axion-monopole interacting
dark energy}

 \author{Emmanuel N. Saridakis}
 \email{msaridak@noa.gr}
 \affiliation{Institute for Astronomy, Astrophysics, Space Applications and
Remote Sensing, National Observatory of Athens, 15236 Penteli, Greece}
 \affiliation{Departamento de Matem\'{a}ticas, Universidad Cat\'{o}lica del 
  Norte, Avda. Angamos 0610, Casilla 1280, Antofagasta, Chile}
 \affiliation{CAS Key Laboratory for Research in Galaxies and Cosmology, 
School 
  of Astronomy and Space Science,
  University of Science and Technology of China, Hefei 230026, China}

\begin{abstract}
We investigate the cosmological dynamics of
axion-monopole interacting dark energy, examining
whether a microscopically motivated dark-sector
interaction can sustain stable late-time acceleration. In particular,
magnetic monopoles constitute dark matter and an
ultralight axion contributes to dark energy, with
their interaction arising from the Witten effect
through field-dependent dyon masses. We formulate
the autonomous system, classify its critical
points and stability, and investigate the evolution
analytically and numerically. We find that
stable de Sitter attractors at finite positive
Hubble parameter require a positive cosmological
constant and correspond to axion-potential minima,
whereas those at maxima are unstable. Nevertheless,
transient axion-driven acceleration remains possible
without an explicit cosmological constant. Furthermore, the
interaction induces matter-era tracking, namely an
attracting slow-drift regime rather than an exact
critical point. The interaction-effective dark-energy
equation of state can cross the phantom divide,
although the physical equation of state remains
non-phantom. Extending the analysis to the full
Bogomol'nyi-Prasad-Sommerfield dyon mass, we find
that the main dynamical features persist, while
the field evolution and interaction strength
depend on the microscopic dyon properties.
Our results demonstrate how a microscopically motivated dark-sector interaction can
give rise to distinctive cosmological evolution and effective dark-energy behavior.
 \end{abstract}


 \maketitle

\section{Introduction}

The standard $\Lambda$CDM cosmological paradigm
provides a successful description of the
evolution of the Universe. Nevertheless,
the nature of dark matter and dark energy
remains unknown, while the cosmological
constant problem and tensions between
different cosmological observations continue
to motivate alternative scenarios \cite{Weinberg:1988cp,Bertone:2004pz}.
Moreover, recent results from the Dark Energy Spectroscopic
Instrument (DESI) have renewed interest in the possibility
of dynamically evolving dark energy \cite{DESI:2025zgx}, while persistent
observational tensions provide further motivation for
extensions of the standard cosmological scenario
\cite{Abdalla:2022yfr,CosmoVerseNetwork:2025alb}.

Theoretical approaches to these questions
follow two broad directions. The first
modifies the gravitational sector, introducing
additional geometrical degrees of freedom
or generalized gravitational interactions
\cite{CANTATA:2021asi,Clifton:2011jh,Capozziello:2011et,Nojiri:2017ncd}.
Representative examples include curvature-based
extensions  \cite{Starobinsky:1980te, Capozziello:2002rd,
DeFelice:2010aj,Nojiri:2005jg, DeFelice:2008wz}, torsion-based
generalizations
\cite{Cai:2015emx, Ferraro:2006jd,
  Linder:2010py, Chen:2010va,Tamanini:2012hg,Kofinas:2014owa,
Kofinas:2014daa,Bahamonde:2015zma, Bahamonde:2016grb}, or non-metricity-based
modifications
\cite{BeltranJimenez:2017tkd,BeltranJimenez:2019tme,Anagnostopoulos:2021ydo,
Heisenberg:2023lru,De:2023xua}.
The second direction retains general relativity and
extends the matter and energy content of the Universe,
giving rise to dynamical dark-energy models such as
quintessence and k-essence \cite{Copeland:2006wr,Cai:2009zp}, or to effective
descriptions
based on thermodynamic considerations  \cite{Saridakis:2026vxo}.

A particularly interesting class is provided by
interacting dark-energy scenarios, in which dark
matter and dark energy exchange energy through a
non-gravitational interaction
\cite{Amendola:1999er,Wang:2016lxa,Wang:2024vmw}.
Such interactions modify the cosmological evolution
of the dark components and can lead to distinctive
effective dark-energy behavior \cite{Amendola:1999er,Wang:2016lxa,Wang:2024vmw,
Amendola:2000uh,Billyard:2000bh,
Zimdahl:2001ar,Wetterich:1994bg,Cai:2004dk,
delCampo:2008sr,Faraoni:2014vra,Koivisto:2015qua,Kumar:2019wfs,Pan:2019gop,
Khoury:2025txd},  while various
interacting scenarios have been confronted with
cosmological observations
\cite{Cai:2009ht,Wei:2010fz,Wang:2015wga,Pan:2019jqh,Yang:2019uzo,Yang:2022csz,
Silva:2025hxw,Bhattacharjee:2025xeb,Yang:2026vkc}.
A central question concerns the physical origin of these interactions. In most
existing models, the energy-transfer term is introduced phenomenologically,
through a prescribed dependence on the dark-sector densities, the Hubble
parameter, or the scalar field. Although useful for investigating
cosmological phenomenology, such prescriptions do not necessarily identify
the microscopic mechanism responsible for the interaction. It is therefore
important to investigate models in which dark-sector energy exchange
follows from an underlying particle-physics construction.

An interesting realization of microscopic dark-sector
interactions involves axions and magnetic monopoles,
both of which can arise in extensions of the Standard
Model. Axions are well-motivated light pseudoscalar
degrees of freedom, originally introduced in connection
with the strong CP problem
\cite{Peccei:1977hh,Peccei:1977ur,Weinberg:1977ma,
Wilczek:1977pj,Kim:1979if,Shifman:1979if,Dine:1981rt,Svrcek:2006yi,
GrillidiCortona:2015jxo,Giannotti:2017hny},
and have subsequently attracted considerable attention
in particle physics and cosmology
\cite{Kim:2008hd,Kawasaki:2013ae,Marsh:2015xka,
DiLuzio:2020wdo}.
Magnetic monopoles, on the other hand, arise as
topologically nontrivial configurations in spontaneously
broken non-Abelian gauge theories
\cite{Dirac:1931kp,tHooft:1974kcl,Polyakov:1974ek,
Julia:1975ff,Prasad:1975kr,Bogomolny:1975de,
Goddard:1976qe,Preskill:1984gd,
 Milton:2006cp,Patrizii:2015uea,Nishino:2019lqm}.
Their dyonic extensions and electric-magnetic duality
have been extensively investigated
\cite{Montonen:1977sn,Weinberg:1983bf,
Seiberg:1994rs,Seiberg:1994aj,Harvey:1996ur,
 Weinberg:2006rq,Balakin:2017nbg,Anosova:2022cjm}.
A particularly interesting connection between axions
and magnetic monopoles arises through the Witten effect
\cite{Witten:1979ey,
Qi:2008ew,Fukuda:2020imw,Choi:2022fgx}, according to which an axion
background induces an electric charge on magnetic
monopoles, giving rise to dyonic states.
The resulting field dependence of the dyon masses
generates an interaction between the axion and monopole
sectors, with potentially important cosmological
consequences
\cite{Fischler:1983sc,Sikivie:1984yz,
Fan:2021ntg,
Sato:2018nqy,Jeong:2023gjc,Banerjee:2024ykz}.

Recently, an axion-monopole interacting dark-sector scenario
was proposed in \cite{Bagherian:2026jxz}, in which magnetic
monopoles constitute dark matter and an ultralight axion
contributes to dark energy. The interaction arises from
the field dependence of dyon masses and can sustain
matter-era axion tracking even when the intrinsic axion
potential is dynamically subdominant. Related studies
have investigated alternative microscopic realizations
of scalar-dependent dark-matter masses
\cite{Delaunay:2026pnu}, as well as dyon formation
and charge relaxation through classical lattice
simulations \cite{Yamada:2026swn}.

Several questions concerning the cosmological dynamics
remain open. These include the existence and stability
of critical points, particularly whether the interaction
can support stable late-time acceleration without a
cosmological constant, and the distinction between
matter-era tracking and exact critical points.
Furthermore, the interaction-effective dark-energy
equation of state may exhibit phantom behavior despite
the canonical nature of the axion. Finally, the quadratic
dyon-mass approximation becomes less reliable at large
effective charges, motivating an investigation using
the full mass expression.

In the present work, we address these questions through
a systematic dynamical analysis of axion-monopole
interacting dark energy. We formulate the cosmological
equations as a closed autonomous system and determine
its critical points and stability. We find that stable
de Sitter attractors at finite positive Hubble parameter
correspond to axion-potential minima and require a
positive cosmological constant, whereas those at maxima
are unstable. Nevertheless, transient axion-driven
acceleration remains possible without an explicit cosmological
constant.

We additionally show that matter-era tracking corresponds
to an attracting slow-drift regime for the axion velocity,
rather than an exact critical point. Through analytical
approximations and numerical integrations, we examine
the field evolution, energy exchange, and transition to
accelerated expansion. We also demonstrate that the
interaction-effective dark-energy equation of state can
cross the phantom divide while the physical equation of
state remains non-phantom. Finally, employing the full
Bogomol'nyi-Prasad-Sommerfield (BPS) dyon mass, we
investigate the robustness of the tracking mechanism
and the sensitivity of the cosmological evolution to
microscopic dyon properties.

The manuscript is organized as follows.
In Sec. \ref{sec:framework} we present the axion-monopole
framework and derive the cosmological
equations. In Sec. \ref{sec:dynamical_system} we construct
the autonomous system, determine its
critical points and stability, and
investigate late-time acceleration.
In Sec. \ref{sec:cosmological_evolution} we examine the matter-era
tracking regime, the numerical evolution
of the dark-energy sector, and the
effects of the full BPS dyon mass.
Finally, Sec. \ref{Conclusions} summarizes our results
and discusses their implications.

\section{Axion-monopole framework and cosmological equations}
\label{sec:framework}

In this section we present the axion-monopole framework
underlying the interacting dark sector and introduce the
relevant particle-physics quantities. We then derive the
corresponding cosmological equations, which provide the
basis for the dynamical-system analysis and numerical
investigation of the subsequent sections.

\subsection{Axion-monopole interactions and dyon masses}
\label{subsec:axion_monopole}

We consider the axion-monopole interacting dark-sector
scenario of \cite{Bagherian:2026jxz}, in which magnetic
monopoles constitute dark matter and an ultralight axion
provides dynamical dark energy. The interaction arises
through the Witten effect, which induces an axion-dependent
electric charge on the monopoles and consequently modifies
their masses, providing a microscopic mechanism for
dark-sector energy exchange.

The underlying dark sector is described by an $SU(2)'$
gauge theory, spontaneously broken to $U(1)'$ by an
adjoint Higgs field with vacuum expectation value $v_D$. Here, the prime
distinguishes the dark gauge symmetry from the corresponding Standard Model gauge
groups.
This symmetry-breaking pattern gives rise to
't Hooft-Polyakov magnetic monopoles, whose mass and
magnetic charge in the Bogomol'nyi-Prasad-Sommerfield
(BPS) limit are
\begin{equation}
m_{\rm M}=\frac{4\pi v_D}{e_D},
\qquad
q_{\rm M}=\frac{4\pi}{e_D},
\label{eq:monopole_properties}
\end{equation}
where $e_D$ is the dark gauge coupling and
we have assumed unit magnetic charge.

The dark gauge sector is coupled to an axion
field $\varphi$ through a topological interaction.
Introducing the dimensionless field
$\theta\equiv\varphi/f_\varphi$, where
$f_\varphi$ is the axion decay constant,
we extend the dark-sector construction
of \cite{Bagherian:2026jxz} to a cosmological
setting including gravity, baryonic matter,
radiation, and a possible cosmological constant.
The corresponding covariant action can be
written schematically as
\begin{eqnarray}
&&
\!\!\!\!\!\!\!\!\!\!\!\!\!\!\!\!\!\!\!\!
S=\int d^4x\sqrt{-g}\Bigg[
\frac{M_{\rm Pl}^2}{2}(R-2\Lambda)
-\frac12(\partial\varphi)^2
-V(\varphi)
\nonumber\\
&&
\ \ \ \ \ \ \ \ \ \ \ \
+\mathcal{L}_D^{(0)}
+\frac{\alpha_D}{8\pi}\theta
F'_{\mu\nu}\widetilde{F}'^{\mu\nu} +\mathcal{L}_b+\mathcal{L}_r
\Bigg],
\label{eq:axion_lagrangian}
\end{eqnarray}
where $M_{\rm Pl}$ is the reduced Planck mass,
$(\partial\varphi)^2\equiv
g^{\mu\nu}\partial_\mu\varphi\partial_\nu\varphi$,
and $\alpha_D=e_D^2/(4\pi)$.
The term $\mathcal{L}_D^{(0)}$ represents the
dark gauge and Higgs dynamics, excluding the
explicit axion coupling, while
$F'_{\mu\nu}$ is the field strength of the
unbroken $U(1)'$ gauge field and
$\widetilde{F}'^{\mu\nu}$ denotes its dual.
The terms $\mathcal{L}_b$ and $\mathcal{L}_r$
describe baryonic matter and radiation,
respectively. The monopoles arise as solitonic
configurations of the dark gauge sector and
are not introduced as additional fundamental
fields in the action.
Finally, $\Lambda$ denotes a possible explicit
constant vacuum-energy contribution.
We retain this term in order to investigate
the cosmological dynamics both in the presence
and in the absence of an explicit cosmological
constant.

We consider a periodic axion potential of the form
\begin{equation}
V(\varphi)=
\Lambda_a^4
\left[1-\cos\left(\frac{\varphi}{f_\varphi}\right)\right],
\label{eq:axion_potential}
\end{equation}
where $\Lambda_a^4=m_\varphi^2f_\varphi^2$
and $m_\varphi$ is the axion mass.
In the construction of \cite{Bagherian:2026jxz},
this potential originates from nonperturbative
effects in the dark gauge sector, with a
characteristic scale exponentially suppressed
as $\Lambda_a^4\sim v_D^4e^{-2\pi/\alpha_D}$.
Such suppression allows the axion to remain
sufficiently light to contribute to the
late-time cosmological evolution.

Through the Witten effect \cite{Witten:1979ey},
the axion background shifts the electric charge
of the magnetic monopoles, giving rise to dyonic
states. For a dyon characterized by an integer
electric-charge quantum number $n$, the electric
charge is
\begin{equation}
q_e=e_D\left(n-\frac{\theta}{2\pi}\right).
\label{eq:dyon_charge}
\end{equation}
Consequently, the dyon mass acquires an explicit
dependence on the axion field. Expanding to
quadratic order in the electric-charge contribution,
we obtain
\begin{equation}
m_D(n,\theta)=m_{\rm M}
\left[
1+\frac{\gamma\alpha_D^2}{2}
\left(n-\frac{\theta}{2\pi}\right)^2
\right],
\label{eq:dyon_mass}
\end{equation}
where $\gamma$ is a dimensionless parameter
of order unity, with $\gamma=1$ in the BPS limit.
This approximation is valid for
$\alpha_D|n-\theta/(2\pi)|\ll1$,
where higher-order contributions can be neglected.
The effects of using the full, unexpanded BPS
mass expression will be examined in Sec. \ref{subsec:bps_dyon_mass}.

The axion dependence of the dyon mass implies
that the dark matter particle masses evolve
together with the scalar field.
To describe a cosmological population of dyons,
we denote by $f_n$ the fraction of particles
occupying each charge state $n$, satisfying
$\sum_n f_n=1$.
Assuming that these fractions remain constant,
the population-averaged dyon mass becomes
\begin{equation}
\overline{m}_D(\theta)
=
\sum_n f_n m_D(n,\theta)
=
m_{\rm M}\mathcal{A}(\theta),
\label{eq:average_dyon_mass}
\end{equation}
where
\begin{equation}
\mathcal{A}(\theta)=
1+\frac{\gamma\alpha_D^2}{2}
\left[
\sigma_n^2+
\left(n_e-\frac{\theta}{2\pi}\right)^2
\right].
\label{eq:mass_function}
\end{equation}
Here we have introduced
\begin{equation}
n_e\equiv\sum_n n f_n,
\qquad
\sigma_n^2\equiv
\sum_n f_n(n-n_e)^2,
\label{eq:charge_moments}
\end{equation}
which represent the mean electric-charge
quantum number and its variance, respectively.
Thus, within the quadratic approximation,
the charge distribution affects the average
dyon mass through these two moments.
The field-dependent part is determined by
$n_e$, while the charge dispersion
$\sigma_n^2$ also modifies the mass normalization.

For a population occupying a single charge
state, $\sigma_n^2=0$, and the mass-modulation
function reduces to
\begin{equation}
\mathcal{A}(\theta)=
1+\frac{\gamma\alpha_D^2}{2}
\left(n_e-\frac{\theta}{2\pi}\right)^2,
\label{eq:mass_function_single}
\end{equation}
recovering the prescription of
\cite{Bagherian:2026jxz}.
We adopt this expression as our reference
case, while retaining the more general
distribution of Eq.~\eqref{eq:mass_function}
when examining the effects of charge dispersion.

The field dependence of the dyon mass leads
to an exchange of energy between dark matter
and the axion sector.
The corresponding coupling function is
defined as
\begin{equation}
\beta(\varphi)
\equiv
\frac{d\ln\mathcal{A}}{d\varphi},
\label{eq:axion_coupling}
\end{equation}
which, using Eq.~\eqref{eq:mass_function},
becomes
\begin{equation}
\beta(\varphi)=
\frac{\gamma\alpha_D^2}
{2\pi f_\varphi\mathcal{A}(\theta)}
\left(\frac{\theta}{2\pi}-n_e\right).
\label{eq:axion_coupling_explicit}
\end{equation}
Hence, the interaction strength is not
an arbitrary constant but is determined
by the axion field and the properties
of the dyon population. Its magnitude
and sign may vary during the cosmological
evolution, while the direction of energy
transfer also depends on the axion velocity,
as discussed in the next subsection.

Finally, the assumption of constant $f_n$
neglects transitions between different
dyon charge states.
Recent classical lattice simulations
suggest that dyons can exchange electric
charge with their environment
\cite{Yamada:2026swn}.
Although these results do not directly
establish the evolution of quantized
dyon populations in the cosmological
setting considered here, they indicate
that fixed charge fractions constitute
a physical approximation.
In the following, we adopt this
approximation and investigate the
resulting cosmological dynamics.

\subsection{Cosmological equations}
\label{subsec:cosmological_equations}

We now investigate the cosmological realization of
the axion-monopole framework described above.
We consider a spatially flat
Friedmann-Lemaître-Robertson-Walker (FLRW)
geometry, with metric
\begin{equation}
ds^2=-dt^2+a^2(t)\delta_{ij}dx^i dx^j,
\label{eq:flrw_metric}
\end{equation}
where $a(t)$ is the scale factor and
$H=\dot a/a$ is the Hubble parameter.
The axion field is assumed to be homogeneous,
$\varphi=\varphi(t)$.
The cosmological energy content consists of
monopole dark matter, the axion field, baryonic
matter, and radiation. We also allow for an
explicit cosmological constant, in order to
investigate the dynamics both in its presence
and in its absence.

As discussed in the previous subsection,
the axion dependence of the dyon masses
is encoded in the function
$\mathcal{A}(\theta)$.
Assuming conservation of the comoving
monopole number, the corresponding
energy density is
\begin{equation}
\rho_{\rm M}
=n_{\rm M}m_{\rm M}\mathcal{A}(\theta)
=n_{{\rm M},0}a^{-3}m_{\rm M}
\mathcal{A}(\theta),
\label{eq:monopole_density}
\end{equation}
where $n_{{\rm M},0}$ is the present monopole
number density.
Thus, unlike conventional cold dark matter,
the monopole energy density does not
necessarily scale as $a^{-3}$, since
the particle masses evolve with the
axion field.

The energy density and pressure of the
canonical axion field are
\begin{align}
\rho_\varphi &=
\frac{1}{2}\dot\varphi^2+V(\varphi),
\label{eq:axion_density}\\
p_\varphi &=
\frac{1}{2}\dot\varphi^2-V(\varphi),
\label{eq:axion_pressure}
\end{align}
where $V(\varphi)$ is given by
Eq.~\eqref{eq:axion_potential}.
The corresponding equation-of-state
parameter is
\begin{equation}
w_\varphi=
\frac{\dot\varphi^2/2-V(\varphi)}
{\dot\varphi^2/2+V(\varphi)}.
\label{eq:axion_eos}
\end{equation}

The two Friedmann equations take the form
\begin{align}
3M_{\rm Pl}^2H^2
&=\rho_{\rm M}+\rho_b+\rho_r
+\rho_\varphi+\rho_\Lambda,
\label{eq:friedmann1}\\
-2M_{\rm Pl}^2\dot H
&=\rho_{\rm M}+\rho_b
+\frac{4}{3}\rho_r+\dot\varphi^2,
\label{eq:friedmann2}
\end{align}
where $\rho_b$ and $\rho_r$ denote the
baryonic matter and radiation energy
densities, respectively, while
$\rho_\Lambda\equiv M_{\rm Pl}^2\Lambda$ is the    constant vacuum
energy density. In our analytical treatment,
radiation is described as a perfect fluid
with $p_r=\rho_r/3$, neglecting the
transition of massive neutrinos to the
nonrelativistic regime.

It is convenient to introduce an effective
dark-energy sector that includes both the
axion field and the possible explicit cosmological
constant. Its energy density and pressure
are defined as
\begin{align}
\rho_{\rm DE}
&\equiv \rho_\varphi+\rho_\Lambda
=\frac{1}{2}\dot\varphi^2
+V(\varphi)+\rho_\Lambda,
\label{eq:dark_energy_density}\\
p_{\rm DE}
&\equiv p_\varphi-\rho_\Lambda
=\frac{1}{2}\dot\varphi^2
-V(\varphi)-\rho_\Lambda.
\label{eq:dark_energy_pressure}
\end{align}
Accordingly, the effective dark-energy
equation-of-state parameter is
\begin{equation}
w_{\rm DE}
\equiv\frac{p_{\rm DE}}{\rho_{\rm DE}}
=
\frac{\dot\varphi^2/2-V(\varphi)-\rho_\Lambda}
{\dot\varphi^2/2+V(\varphi)+\rho_\Lambda}.
\label{eq:dark_energy_eos}
\end{equation}
This definition applies to both cases
considered in the following. In the
presence of an explicit cosmological
constant, dark energy receives contributions
from both the axion and the constant
vacuum energy. For $\Lambda=0$,
the effective dark-energy sector consists
entirely of the axion field, and therefore
$\rho_{\rm DE}=\rho_\varphi$ and
$w_{\rm DE}=w_\varphi$.
In either case, the first Friedmann
equation can be expressed as
$3M_{\rm Pl}^2H^2=
\rho_{\rm M}+\rho_b+\rho_r+\rho_{\rm DE}$.

The characteristic feature of the present
model is the interaction between the axion
and monopole sectors.
Differentiating Eq.~\eqref{eq:monopole_density}
and using the conservation of the monopole
number, we obtain
\begin{equation}
\dot\rho_{\rm M}+3H\rho_{\rm M}=Q,
\label{eq:monopole_conservation}
\end{equation}
where the energy-transfer term is
\begin{equation}
Q=
\rho_{\rm M}
\frac{d\ln\mathcal{A}}{d\varphi}
\dot\varphi
=\beta(\varphi)\rho_{\rm M}\dot\varphi,
\label{eq:interaction_Q}
\end{equation}
with $\beta(\varphi)$ given by
Eq.~\eqref{eq:axion_coupling_explicit}.
The axion sector consequently satisfies
\begin{equation}
\dot\rho_\varphi+
3H(\rho_\varphi+p_\varphi)=-Q.
\label{eq:axion_conservation}
\end{equation}
With this convention, $Q>0$ corresponds
to energy transfer from the axion to
the monopole sector, while $Q<0$
describes the opposite direction.
The sign and magnitude of the interaction
are therefore determined by the axion
evolution and the populated dyon
charge states.

Since the cosmological constant is
separately conserved, $\dot\rho_\Lambda=0$,
the effective dark-energy sector obeys
\begin{equation}
\dot\rho_{\rm DE}
+3H(\rho_{\rm DE}+p_{\rm DE})=-Q.
\label{eq:dark_energy_conservation}
\end{equation}
The energy-transfer term can be absorbed
into an interaction-effective dark-energy
equation-of-state parameter by rewriting
Eq.~\eqref{eq:dark_energy_conservation} as
$\dot\rho_{\rm DE}
+3H(1+w_{\rm DE}^{\rm eff})\rho_{\rm DE}=0$.
Comparing the two forms, we obtain
\begin{equation}
w_{\rm DE}^{\rm eff}
=w_{\rm DE}+\frac{Q}{3H\rho_{\rm DE}}.
\label{eq:wDE_effective}
\end{equation}
Unlike the physical dark-energy equation
of state $w_{\rm DE}$, the interaction-effective
quantity $w_{\rm DE}^{\rm eff}$ can fall
below the phantom divide $-1$ due to   the energy exchange between the dark
sectors, without requiring a phantom axion.
Thus, although the effective dark-energy
and monopole components are not separately
conserved, their combined energy-momentum
tensor is conserved. Importantly, the
interaction acts only on the axion
component, while the cosmological
constant remains unaffected.

Substituting Eqs.~\eqref{eq:axion_density}
and \eqref{eq:axion_pressure} into
Eq.~\eqref{eq:axion_conservation},
we obtain the axion equation of motion
\begin{equation}
\ddot\varphi+3H\dot\varphi
+\frac{dV}{d\varphi}
=-\beta(\varphi)\rho_{\rm M}.
\label{eq:axion_eom}
\end{equation}
Expressing this equation in terms of
the dimensionless field
$\theta=\varphi/f_\varphi$ yields
\begin{equation}
\ddot\theta+3H\dot\theta
+m_\varphi^2\sin\theta
+\frac{\rho_{\rm M}}{f_\varphi^2}
\frac{\mathcal{A}_{,\theta}}
{\mathcal{A}}=0,
\label{eq:theta_eom}
\end{equation}
where $\mathcal{A}_{,\theta}\equiv
d\mathcal{A}/d\theta$.
The last term represents the force
exerted on the axion by the monopole
population. Since its magnitude depends
on the monopole density, the interaction
introduces a density-dependent
modification of the axion dynamics,
absent in uncoupled quintessence models.

In order to study the cosmological
evolution, we introduce the e-folding
number $N\equiv\ln a$, with $a=1$
at the present epoch. We further define
\begin{equation}
E\equiv\frac{H}{H_0},
\qquad
\mu\equiv\frac{m_\varphi}{H_0},
\qquad
\tilde f\equiv\frac{f_\varphi}{M_{\rm Pl}},
\label{eq:background_parameters}
\end{equation}
where $H_0$ denotes the present Hubble
parameter. In the following, a prime
denotes differentiation with respect
to $N$.

Using these definitions, and introducing the density parameters, the first
Friedmann equation becomes
\begin{eqnarray}
&&
\!\!\!\!\!\!\!\!\!\!\!\!\!\!\!\!
\left[3-\frac{\tilde f^2}{2}
(\theta')^2\right]E^2
=
\mu^2\tilde f^2(1-\cos\theta)
+3\Omega_{\Lambda0}
\nonumber\\
&&
 \!\!\!\!\!\!\!\!
+3\Omega_{{\rm M},*}e^{-3N}
\mathcal{A}(\theta)
+3\Omega_{b0}e^{-3N}
+3\Omega_{r0}e^{-4N},
\label{eq:normalized_friedmann}
\end{eqnarray}
where
\begin{equation}
\Omega_{{\rm M},*}
\equiv
\frac{n_{{\rm M},0}m_{\rm M}}
{3M_{\rm Pl}^2H_0^2},
\qquad
\Omega_{\Lambda0}\equiv
\frac{\rho_\Lambda}
{3M_{\rm Pl}^2H_0^2}.
\label{eq:monopole_normalization}
\end{equation}
As usual, $\Omega_{b0}$ and
$\Omega_{r0}$ denote the present
baryonic matter and radiation density
parameters, respectively.

We note that $\Omega_{{\rm M},*}$ is
associated with the conserved monopole
abundance and does not coincide with
the physical monopole density parameter.
The latter is given by
\begin{equation}
\Omega_{\rm M}(N)=
\frac{\Omega_{{\rm M},*}e^{-3N}
\mathcal{A}(\theta)}{E^2},
\label{eq:monopole_omega}
\end{equation}
and hence
$\Omega_{{\rm M},0}
=\Omega_{{\rm M},*}\mathcal{A}(\theta_0)$,
where $\theta_0$ is the present
axion field value. This distinction
is important when comparing solutions
corresponding to different dyon-mass
prescriptions.

Similarly, the axion density parameter
can be expressed as
\begin{equation}
\Omega_\varphi(N)=
\frac{\tilde f^2}{6}(\theta')^2
+\frac{\mu^2\tilde f^2}{3E^2}
(1-\cos\theta).
\label{eq:axion_omega}
\end{equation}
The effective dark-energy density
parameter is then
\begin{equation}
\Omega_{\rm DE}(N)
=\Omega_\varphi(N)
+\frac{\Omega_{\Lambda0}}{E^2}.
\label{eq:dark_energy_omega}
\end{equation}
Here, $\Omega_{\Lambda0}$ is normalized
with respect to the present critical
density and remains constant, whereas
its contribution to the instantaneous
density budget is $\Omega_{\Lambda0}/E^2$.
The Friedmann constraint therefore gives
\begin{equation}
\Omega_{\rm M}
+\Omega_b+\Omega_r+\Omega_{\rm DE}=1,
\label{eq:omega_constraint}
\end{equation}
where $\Omega_b=\Omega_{b0}e^{-3N}/E^2$
and $\Omega_r=\Omega_{r0}e^{-4N}/E^2$.
For $\Omega_{\Lambda0}=0$, we simply have
$\Omega_{\rm DE}=\Omega_\varphi$.

The axion evolution equation becomes
\begin{equation}
 \theta''
+\left(3+\frac{E'}{E}\right)\theta'
+\frac{\mu^2}{E^2}\sin\theta
+\frac{3\Omega_{{\rm M},*}e^{-3N}}
{\tilde f^2E^2}
\mathcal{A}_{,\theta}=0,
\label{eq:normalized_axion}
\end{equation}
while the second Friedmann equation gives
\begin{eqnarray}
&&
\!\!\!\!\!\!\!\!\!\!\!\!\!\!\!\!\!\!\!\!\!\!
\frac{E'}{E}
=-\frac{3}{2E^2}
\Big[
\Omega_{{\rm M},*}e^{-3N}
\mathcal{A}(\theta)
+\Omega_{b0}e^{-3N}
\nonumber\\
&& \ \ \ \ \ \ \,
+\frac{4}{3}\Omega_{r0}e^{-4N}
\Big]-\frac{\tilde f^2}{2}(\theta')^2.
\label{eq:normalized_hubble}
\end{eqnarray}
Additionally, at the present epoch, the normalization
$E(0)=1$ imposes the constraint
\begin{align}
1={}&
\Omega_{{\rm M},*}\mathcal{A}(\theta_0)
+\Omega_{b0}+\Omega_{r0}
+\Omega_{\Lambda0}
\nonumber\\
&+\frac{\tilde f^2}{6}(\theta'_0)^2
+\frac{\mu^2\tilde f^2}{3}
(1-\cos\theta_0).
\label{eq:present_friedmann_constraint}
\end{align}
This relation fixes one of the
background parameters once the
remaining parameters and the present
axion configuration are specified.
Equivalently, it corresponds to
$\Omega_{{\rm M},0}+\Omega_{b0}
+\Omega_{r0}+\Omega_{{\rm DE},0}=1$.

The above equations determine the
background evolution of the interacting
axion-monopole cosmology.
The deceleration parameter is given by
\begin{equation}
q=-1-\frac{E'}{E},
\label{eq:effective_eos_deceleration}
\end{equation}
and thus accelerated expansion occurs for $q<0$.

We also note that the axion-dependent
contribution to the monopole mass is
already incorporated in $\rho_{\rm M}$
through $\mathcal{A}(\theta)$ and
should not be introduced as an
additional independent energy-density
component in the Friedmann equations.
Moreover, for a canonical axion with
nonnegative potential and
$\rho_\Lambda\geq0$, the physical
dark-energy sector satisfies
$w_{\rm DE}\geq-1$ whenever
$\rho_{\rm DE}>0$.
Nevertheless, the interaction can
produce an apparent phantom equation
of state when the expansion history
is interpreted under the assumption
of separately conserved dark matter.

\section{Dynamical-system analysis}
\label{sec:dynamical_system}

In   this section  we investigate the phase-space dynamics of the
interacting axion-monopole cosmology, determining
its critical points and their stability  \cite{Copeland:1997et,Bahamonde:2017ize}.
Particular attention is devoted to the
matter-dominated regime and the existence of
stable late-time accelerated solutions.

\subsection{Autonomous variables and dynamical equations}
\label{subsec:autonomous_system}
In order to cast the cosmological equations into
an autonomous form, we introduce the dimensionless
variables
\begin{align}
x &\equiv
\frac{\dot\varphi}{\sqrt{6}M_{\rm Pl}H}
=\frac{\tilde f}{\sqrt{6}}\theta',
\nonumber\\
y &\equiv \frac{H_0}{H}=\frac{1}{E},
\nonumber\\
v &\equiv
\frac{\rho_{\rm M}}{3M_{\rm Pl}^2H^2},
\nonumber\\
r &\equiv
\frac{\rho_r}{3M_{\rm Pl}^2H^2},
\nonumber\\
u &\equiv \theta,
\label{eq:autonomous_variables}
\end{align}
where a prime denotes differentiation with
respect to $N=\ln a$.
The variable $x$ describes the axion kinetic
contribution, while $v$ and $r$ represent
the monopole and radiation density parameters.
The variables $y$ and $u$ characterize the
evolution of the Hubble parameter and the
axion field, respectively.
Retaining $u$ as an independent variable
is particularly useful for the periodic
potential of Eq.~\eqref{eq:axion_potential},
since the usual potential-slope variable
$-M_{\rm Pl}V_{,\varphi}/V$ becomes singular
at its minima. Our choice avoids this
difficulty and preserves the exact field
dependence of both the potential and the
monopole mass.
The axion kinetic and potential density
parameters are
\begin{equation}
\Omega_{\rm kin}=x^2,
\qquad
\Omega_V=
\frac{\mu^2\tilde f^2}{3}
y^2(1-\cos u),
\label{eq:axion_density_variables}
\end{equation}
and hence
\begin{equation}
\Omega_\varphi=x^2+\Omega_V.
\label{eq:omega_axion}
\end{equation}
The instantaneous density parameter
associated with the cosmological constant is
\begin{equation}
\Omega_\Lambda=\Omega_{\Lambda0} y^2,
\label{eq:omega_lambda_variable}
\end{equation}
where $\Omega_{\Lambda0}$ on the right-hand
side is the constant parameter defined in
Eq.~\eqref{eq:monopole_normalization}.
According to the effective dark-energy
definition introduced in Sec. \ref{subsec:cosmological_equations},
we therefore obtain
\begin{equation}
\Omega_{\rm DE}
=\Omega_\varphi+\Omega_{\Lambda0} y^2
=x^2+\Omega_V+\Omega_{\Lambda0} y^2.
\label{eq:omega_de_autonomous}
\end{equation}
For $\Omega_{\Lambda0}=0$, the dark-energy
sector consists entirely of the axion,
and $\Omega_{\rm DE}=\Omega_\varphi$.
The first Friedmann equation
\eqref{eq:friedmann1} provides the constraint
\begin{equation}
1=x^2+\Omega_V+v+b+r+\Omega_{\Lambda0} y^2,
\label{eq:autonomous_constraint}
\end{equation}
where $b\equiv\rho_b/(3M_{\rm Pl}^2H^2)$
is the baryonic matter density parameter.
We can thus eliminate $b$ through
\begin{equation}
b=
1-x^2-v-r
-\frac{\mu^2\tilde f^2}{3}
y^2(1-\cos u)
-\Omega_{\Lambda0} y^2.
\label{eq:baryon_constraint}
\end{equation}
Defining the total matter density parameter
as $\Omega_m\equiv b+v$, including both
baryons and monopoles, the constraint
can also be written as
$\Omega_m+r+\Omega_{\rm DE}=1$.
The axion-monopole interaction enters
through the field-dependent coupling.
For convenience, we introduce
\begin{equation}
\mathcal{B}(u)
\equiv M_{\rm Pl}\beta(\varphi)
=\frac{1}{\tilde f}
\frac{\mathcal{A}_{,u}}{\mathcal{A}},
\label{eq:dimensionless_coupling}
\end{equation}
where $\mathcal{A}(u)$ is given by
Eq.~\eqref{eq:mass_function}.
Explicitly,
\begin{equation}
\mathcal{B}(u)=
\frac{\gamma\alpha_D^2}
{2\pi\tilde f\,\mathcal{A}(u)}
\left(\frac{u}{2\pi}-n_e\right).
\label{eq:coupling_explicit_autonomous}
\end{equation}
Thus, the charge distribution affects
the coupling through its mean $n_e$
and the mass function $\mathcal{A}(u)$,
which also depends on the variance
$\sigma_n^2$. The single-charge case
is recovered for $\sigma_n^2=0$.
We further define
\begin{equation}
\varepsilon\equiv
-\frac{\dot H}{H^2}
=-\frac{H'}{H}.
\label{eq:epsilon_definition}
\end{equation}
Using the second Friedmann equation
\eqref{eq:friedmann2}, we obtain
\begin{equation}
\varepsilon=
3x^2+\frac{3}{2}(v+b)+2r,
\label{eq:epsilon_friedmann}
\end{equation}
and eliminating $b$ through
Eq.~\eqref{eq:baryon_constraint},
this becomes
\begin{equation}
\varepsilon=
\frac{3}{2}
\left[
1+x^2+\frac{r}{3}
-\frac{\mu^2\tilde f^2}{3}
y^2(1-\cos u)
-\Omega_{\Lambda0} y^2
\right].
\label{eq:epsilon_autonomous}
\end{equation}
Differentiating the variables of
Eq.~\eqref{eq:autonomous_variables}
with respect to $N$, and using the
cosmological equations of Sec.   \ref{sec:framework},
we obtain
\begin{align}
x'={}&
(\varepsilon-3)x
-\frac{\mu^2\tilde f}{\sqrt{6}}
y^2\sin u
-\sqrt{\frac{3}{2}}\,
\mathcal{B}(u)v,
\label{eq:autonomous_x}\\
y'={}&
\varepsilon y,
\label{eq:autonomous_y}\\
v'={}&
v\left[
2\varepsilon-3
+\sqrt{6}\mathcal{B}(u)x
\right],
\label{eq:autonomous_z}\\
r'={}&
(2\varepsilon-4)r,
\label{eq:autonomous_r}\\
u'={}&
\frac{\sqrt{6}}{\tilde f}x.
\label{eq:autonomous_u}
\end{align}
Equations
\eqref{eq:autonomous_x}-\eqref{eq:autonomous_u},
together with
Eqs.~\eqref{eq:coupling_explicit_autonomous}
and \eqref{eq:epsilon_autonomous},
constitute a closed five-dimensional
autonomous system.
The interaction appears explicitly
in the equations for $x$ and $v$.
The last term of Eq.~\eqref{eq:autonomous_x}
describes the monopole-induced force
on the axion, while the coupling term
in Eq.~\eqref{eq:autonomous_z} modifies
the standard dilution of the monopole
energy density. When the coupling
vanishes identically, the system
reduces to uncoupled axion quintessence
and conventional cold dark matter.
As a consistency check, differentiating
Eq.~\eqref{eq:baryon_constraint} and
using the autonomous equations yields
\begin{equation}
b'=(2\varepsilon-3)b,
\label{eq:autonomous_baryons}
\end{equation}
which reproduces the conservation
equation for pressureless baryonic
matter. Therefore, the Friedmann
constraint is preserved, and $b$
need not be introduced as an
independent dynamical variable.
The physically allowed phase space
is determined by
\begin{equation}
y>0,\qquad v\geq0,\qquad r\geq0,
\qquad b\geq0,
\label{eq:physical_phase_space}
\end{equation}
with $b$ given by
Eq.~\eqref{eq:baryon_constraint}.
We consider expanding solutions
with $H>0$ and $\Omega_{\Lambda0}\geq0$.
The Friedmann constraint implies
$|x|\leq1$ and, for
$\Omega_{\Lambda0}>0$,
$y\leq\Omega_{\Lambda0}^{-1/2}$.
The axion variable $u$ may take any
real value. Although its intrinsic
potential is periodic, the mass
function $\mathcal{A}(u)$ is not
periodic on a fixed dyon-charge
branch. Hence, the identification
$u\sim u+2\pi$ requires a corresponding
relabeling of the dyon charge states.
The cosmological observables can be
expressed directly in terms of the
autonomous variables. In particular,
the  deceleration
parameter is
\begin{equation}
q=-1+\varepsilon,
\label{eq:autonomous_observables}
\end{equation}
and thus
accelerated expansion
  corresponds to
$\varepsilon<1$. Moreover,
  the intrinsic axion
equation-of-state parameter is
\begin{equation}
w_\varphi=
\frac{x^2-\Omega_V}
{x^2+\Omega_V},
\label{eq:autonomous_axion_eos}
\end{equation}
provided that $\Omega_\varphi\neq0$.
Similarly, the equation-of-state
parameter of the physical dark-energy
sector defined in (\ref{eq:dark_energy_eos}) becomes
\begin{equation}
w_{\rm DE}=
\frac{x^2-\Omega_V-\Omega_{\Lambda0} y^2}
{x^2+\Omega_V+\Omega_{\Lambda0} y^2},
\label{eq:autonomous_dark_energy_eos}
\end{equation}
provided that $\Omega_{\rm DE}\neq0$.
Additionally, the interaction-effective
dark-energy equation-of-state parameter (\ref{eq:wDE_effective})
 takes the form
\begin{equation}
w_{\rm DE}^{\rm eff}=
\frac{x^2-\Omega_V-\Omega_{\Lambda0} y^2
+\sqrt{\frac{2}{3}}\,\mathcal{B}(u)xv}
{x^2+\Omega_V+\Omega_{\Lambda0} y^2},
\label{eq:autonomous_wDE_effective}
\end{equation}
where $\mathcal{B}(u)$ is the coupling function
defined in Eq.~(\ref{eq:dimensionless_coupling}). This expression is defined for
$\Omega_{\rm DE}\neq 0$.
While the physical equation of state
satisfies $w_{\rm DE}\geq-1$, the
interaction-effective quantity can
cross the phantom divide due to the
energy exchange between the axion
and monopole sectors.

Finally, retaining $u$ as an independent variable
avoids singularities associated with the minima
of the periodic axion potential.
However, for $\Omega_{\Lambda0}=0$, the asymptotic
limit $H\rightarrow0$ corresponds to
$y\rightarrow\infty$ and lies at the boundary
of the noncompact phase space.
Such configurations require a separate
asymptotic analysis.

\subsection{Critical points and stability}
\label{subsec:critical_points}

We proceed to determine the critical points of the
autonomous system
\eqref{eq:autonomous_x}-\eqref{eq:autonomous_u}
and investigate their stability by examining the
eigenvalues of the corresponding Jacobian matrix.
Throughout the analysis, we consider
$\mu>0$, $\tilde f>0$, $\gamma\alpha_D^2>0$,
$\Omega_{\Lambda0}\geq0$, and nonnegative energy
densities.

From Eq.~\eqref{eq:autonomous_u}, every critical
point with finite phase-space coordinates satisfies
$x_c=0$, while Eq.~\eqref{eq:autonomous_y}
implies $\varepsilon_c y_c=0$.
We therefore distinguish between critical points
with $y_c>0$, corresponding to a finite positive
Hubble parameter, and those located at the boundary
$y_c=0$, which describes the asymptotic regime
$H/H_0\to\infty$.

For $y_c>0$, we necessarily have
$\varepsilon_c=0$.
Since Eq.~\eqref{eq:epsilon_friedmann} gives
$\varepsilon=3x^2+3(v+b)/2+2r\geq0$,
this condition requires
$x_c=v_c=b_c=r_c=0$.
The remaining critical-point equation reduces
to $\sin u_c=0$, implying that all finite-Hubble
critical points correspond to extrema of the
axion potential. These solutions form two
de Sitter families, associated with the
potential minima and maxima.

At the high-Hubble boundary $y_c=0$, we obtain
$\varepsilon_c=(3+r_c)/2$.
The remaining conditions become
$\mathcal{B}(u_c)v_c=0$, $v_cr_c=0$, and
$r_c(r_c-1)=0$.
They give rise to a radiation-dominated family
and two matter-dominated families, distinguished
by the presence or absence of monopoles.

The resulting critical points are summarized
in Table~\ref{tab:critical_points}, where we
have introduced
$L\equiv\Omega_{\Lambda0}+2\mu^2\tilde f^2/3$.
We also present the total matter and effective
dark-energy density parameters, defined in the
previous subsection as
$\Omega_m=b+v$ and
$\Omega_{\rm DE}=x^2+\Omega_V+\Omega_{\Lambda0} y^2$.
In particular, both de Sitter families satisfy
$\Omega_{\rm DE}=1$, whereas the radiation- and
matter-dominated boundary points have
$\Omega_{\rm DE}=0$.
Consequently, the physical and interaction-effective
dark-energy equation-of-state parameters are
well defined at the de Sitter points, but
undefined at the radiation- and matter-dominated
critical points.

\begin{table*}[t]
\caption{Critical points of the autonomous system,
their existence conditions, physical and
interaction-effective dark-energy equation-of-state
parameters, radiation, matter and dark-energy
density parameters, and stability.
The coordinates are ordered as
$(x_c,y_c,v_c,r_c,u_c)$, with $k\in\mathbb{Z}$ and
$L\equiv\Omega_{\Lambda0}+2\mu^2\tilde f^2/3$.
The parameter $u_c$ is arbitrary for the radiation-
and baryon-dominated families.
The total matter density parameter
$\Omega_m=b+v$ includes baryons and monopoles,
while $\Omega_r=r$ and
$\Omega_{\rm DE}=\Omega_\varphi+\Omega_{\Lambda0} y^2$.
The quantities $w_{\rm DE}$ and
$w_{\rm DE}^{\rm eff}$ are undefined when
$\Omega_{\rm DE}=0$.
For $\mathcal{M}_{\rm M}$, the monopole and baryon
fractions are $v_c$ and $1-v_c$, respectively.
The last three families lie on the high-Hubble
boundary $y_c=0$.}
\label{tab:critical_points}
\small
\begin{ruledtabular}
\begin{tabular}{lcccccccc}
Critical point &
$(x_c,y_c,v_c,r_c,u_c)$ &
Existence &
$w_{\rm DE}$ &
$w_{\rm DE}^{\rm eff}$ &
$\Omega_r$ &
$\Omega_m$ &
$\Omega_{\rm DE}$ &
Stability \\
$\mathcal{D}_{\rm min}^{(k)}$ &
$(0,\Omega_{\Lambda0}^{-1/2},0,0,2\pi k)$ &
$\Omega_{\Lambda0}>0$ &
$-1$ &
$-1$ &
$0$ &
$0$ &
$1$ &
Stable \\
$\mathcal{D}_{\rm max}^{(k)}$ &
$(0,L^{-1/2},0,0,(2k+1)\pi)$ &
$\Omega_{\Lambda0}\geq0$ &
$-1$ &
$-1$ &
$0$ &
$0$ &
$1$ &
Saddle \\
$\mathcal{R}(u_c)$ &
$(0,0,0,1,u_c)$ &
All &
\text{Undefined} &
\text{Undefined} &
$1$ &
$0$ &
$0$ &
Saddle \\
$\mathcal{M}_b(u_c)$ &
$(0,0,0,0,u_c)$ &
All &
\text{Undefined} &
\text{Undefined} &
$0$ &
$1$ &
$0$ &
Saddle \\
$\mathcal{M}_{\rm M}(v_c)$ &
$(0,0,v_c,0,2\pi n_e)$ &
$0<v_c\leq1$ &
\text{Undefined} &
\text{Undefined} &
$0$ &
$1$ &
$0$ &
Saddle
\end{tabular}
\end{ruledtabular}
\end{table*}

We first examine the de Sitter solutions.
At the potential minima
$\mathcal{D}_{\rm min}^{(k)}$, which exist
for $\Omega_{\Lambda0}>0$, the eigenvalues
are $-3,-3,-4$, together with
\begin{equation}
\xi_\pm=
-\frac{3}{2}\pm
\frac{1}{2}
\sqrt{9-\frac{4\mu^2}{\Omega_{\Lambda0}}}.
\label{eq:min_desitter_eigenvalues}
\end{equation}
All eigenvalues have negative real parts
for $\mu>0$ and $\Omega_{\Lambda0}>0$.
These points are therefore stable late-time
attractors,
corresponding to a dark-energy-dominated Universe with $q=-1$ and
$w_{\rm DE}=w_{\rm DE}^{\rm eff}=-1$.
Depending on the ratio $\mu^2/\Omega_{\Lambda0}$,
the approach to the attractor is monotonic
or oscillatory.

At the potential maxima
$\mathcal{D}_{\rm max}^{(k)}$, the eigenvalues
again include $-3,-3,-4$, while the
remaining pair is
\begin{equation}
\xi_\pm=
-\frac{3}{2}\pm
\frac{1}{2}
\sqrt{9+\frac{4\mu^2}{L}}.
\label{eq:max_desitter_eigenvalues}
\end{equation}
Since one eigenvalue is always positive,
these solutions are saddle points.
They are nevertheless characterized by
$q=-1$, $\Omega_{\rm DE}=1$, and
$w_{\rm DE}=w_{\rm DE}^{\rm eff}=-1$,
and exist even for $\Omega_{\Lambda0}=0$.
In the latter case, accelerated expansion
is supported entirely by the axion potential,
although the corresponding configuration
is unstable.

We mention here that although both de Sitter families satisfy
$\Omega_{\rm DE}=1$ and
$w_{\rm DE}=w_{\rm DE}^{\rm eff}=-1$,
they correspond to physically distinct
configurations. At the potential minima,
the expansion is driven solely by the
cosmological constant, with
$H_c^2=H_0^2\Omega_{\Lambda0}$, and the
solutions are stable late-time attractors.
At the potential maxima, the axion
potential provides an additional
contribution, yielding $H_c^2=H_0^2L$,
while the corresponding solutions
are unstable saddle points.

The locations and eigenvalues of both
de Sitter families are independent of the
axion-monopole coupling parameters.
Indeed, the monopole density vanishes at
these points, and the interaction does not
affect their linear stability.
Consequently, the energy-transfer term
vanishes, and the physical and
interaction-effective dark-energy equations
of state coincide.
Nevertheless, the coupling influences the
preceding evolution and may affect which
potential minimum is eventually reached.

We now turn to the critical points on the
high-Hubble boundary.
The radiation-dominated family
$\mathcal{R}(u_c)$ has eigenvalues
$\{2,1,1,-1,0\}$, while the
baryon-dominated family
$\mathcal{M}_b(u_c)$ has
$\{3/2,-3/2,-1,0,0\}$.
Both families are nonhyperbolic and
possess unstable directions.
The zero eigenvalue of the radiation
family reflects the freedom in $u_c$,
while the baryon-dominated family has
an additional marginal direction at
linear order.

These solutions represent the familiar
radiation- and matter-dominated regimes,
with $q=1$ and $q=1/2$, respectively.
In both cases $\Omega_{\rm DE}=0$,
and hence neither $w_{\rm DE}$ nor
$w_{\rm DE}^{\rm eff}$ is defined
at the exact critical point.

Of particular interest is the
matter-dominated family
$\mathcal{M}_{\rm M}(v_c)$, which contains
a nonvanishing monopole contribution.
Its existence requires
$\mathcal{B}(u_c)=0$, yielding
$u_c=2\pi n_e$ for the quadratic
mass-modulation function.
Thus, the axion is located at the minimum
of the monopole mass function, where the
interaction-induced force vanishes despite
the presence of monopoles.
The corresponding density fractions satisfy
$\Omega_m=1$ and $\Omega_{\rm DE}=0$,
with $q=1/2$, and with $v_c$ and
$1-v_c$ representing the monopole and
baryon contributions.

Defining
\begin{equation}
\mathcal{A}_c
\equiv\mathcal{A}(2\pi n_e)
=1+\frac{\gamma\alpha_D^2}{2}\sigma_n^2,
\label{eq:critical_mass_value}
\end{equation}
the eigenvalues of the Jacobian matrix are
$3/2,-1,0$, together with
\begin{equation}
\xi_\pm=
-\frac{3}{4}\pm
\frac{1}{4}\sqrt{9-\Delta},
\label{eq:monopole_matter_eigenvalues}
\end{equation}
where
$
\Delta=
\frac{12\gamma\alpha_D^2v_c}
{\pi^2\tilde f^2\mathcal{A}_c}>0$.
The last two eigenvalues have negative real
parts, showing that small axion perturbations
around the minimum of the monopole mass
function are damped.
However, the positive eigenvalue $3/2$
implies that this family is unstable towards
the future. It can therefore describe a
matter-dominated stage of cosmic evolution,
but not a late-time attractor.
The vanishing eigenvalue is associated with
the continuous parameter $v_c$.

However, the conservation of monopole and
baryon comoving abundances imposes an
additional restriction.
Using Eqs.~\eqref{eq:autonomous_z}
and \eqref{eq:autonomous_baryons}, we find
that
\begin{equation}
\mathcal{I}
\equiv
\frac{v}{b\mathcal{A}(u)}
=
\frac{\Omega_{{\rm M},*}}{\Omega_{b0}}
\label{eq:abundance_invariant}
\end{equation}
is constant along trajectories with
nonvanishing baryon and monopole densities.
Consequently, for fixed conserved
abundances, the monopole fraction at
the matter-dominated critical point is
\begin{equation}
v_c=
\frac{\Omega_{{\rm M},*}\mathcal{A}_c}
{\Omega_{b0}+
\Omega_{{\rm M},*}\mathcal{A}_c}.
\label{eq:physical_matter_fraction}
\end{equation}
Thus, although the autonomous system admits
a continuous family parametrized by $v_c$,
a fixed ratio of conserved abundances
selects a particular member of this family.
The corresponding zero eigenvalue is
therefore related to variations of the
abundance ratio, rather than to an
additional propagating degree of freedom.
The baryon-free endpoint $v_c=1$
is understood separately as the limiting
case $\Omega_{b0}=0$.

An important distinction should be made
between these matter-dominated critical
points and the Hubble-tracking regime
discussed in \cite{Bagherian:2026jxz}.
At the critical points, the axion field
is stationary, with $x_c=u'_c=0$.
In contrast, Hubble tracking involves
a slowly evolving field, driven by the
density-dependent monopole interaction.
Such configurations therefore correspond
to time-dependent, quasi-stationary
trajectories rather than exact critical
points of the autonomous system.
In particular, for a nonvanishing monopole
density, an exact matter-dominated
critical point requires
$\mathcal{B}(u_c)=0$, which is not
generically satisfied when the axion
is displaced from $u=2\pi n_e$.

The above results establish that the
only stable accelerated critical points
at finite positive Hubble parameter
are the de Sitter minima
$\mathcal{D}_{\rm min}^{(k)}$, which
require $\Omega_{\Lambda0}>0$.
When $\Omega_{\Lambda0}=0$, these points
are displaced to the asymptotic boundary
$y\to\infty$, while the remaining
finite-Hubble de Sitter solutions are
unstable. Hence, the axion-monopole
interaction does not generate an
additional stable accelerated critical
point at finite positive $H$ in the
absence of a cosmological constant.
This conclusion does not exclude
transient acceleration driven by the
evolving axion field.
Furthermore, the limit $H\to0$,
corresponding to $y\to\infty$,
is not covered by the finite critical-point
analysis and requires a separate
asymptotic investigation.
We examine these possibilities in the
following subsection.

\subsection{Late-time attractors and accelerated solutions}
\label{subsec:late_time_attractors}

We now examine the late-time behavior of the
axion-monopole cosmology, focusing on the
existence of accelerated solutions and their
dependence on an explicit cosmological constant.
We first consider $\Omega_{\Lambda0}>0$.
As established in the previous subsection,
the stable critical points correspond to
the minima of the axion potential,
$\mathcal{D}_{\rm min}^{(k)}$, with $u_c=2\pi k$.
At these points, the matter and radiation
densities, as well as the axion kinetic
energy, vanish, and the Hubble parameter
approaches
\begin{equation}
H_{\rm dS}=H_0\sqrt{\Omega_{\Lambda0}}
=\sqrt{\frac{\rho_\Lambda}{3M_{\rm Pl}^2}}.
\label{eq:late_desitter_hubble}
\end{equation}
The corresponding asymptotic behavior is
\begin{equation}
\Omega_{\rm DE}\to1,\qquad
w_{\rm DE}\to-1,\qquad
w_{\rm DE}^{\rm eff}\to-1,\qquad q\to-1.
\label{eq:late_desitter_limit}
\end{equation}
In particular, $\Omega_\varphi\to0$ and
$\Omega_\Lambda\to1$, showing that the
asymptotic acceleration is supported by
the constant vacuum-energy contribution.

This result can also be understood directly
from the Friedmann equations. Indeed,
Eq.~\eqref{eq:friedmann2} gives
\begin{equation}
\dot H=
-\frac{1}{2M_{\rm Pl}^2}
\left(
\rho_{\rm M}+\rho_b+
\frac{4}{3}\rho_r+\dot\varphi^2
\right)\leq0,
\label{eq:late_hubble_monotonicity}
\end{equation}
while the first Friedmann equation implies
$H^2\geq\rho_\Lambda/(3M_{\rm Pl}^2)$.
Thus, for expanding solutions with
$\Omega_{\Lambda0}>0$, the Hubble parameter
is monotonically decreasing and bounded
from below. As the matter components
dilute, the axion approaches an extremum
of its potential. Since the maxima are
unstable, trajectories settling within
a potential well approach the corresponding
de Sitter minimum.

The approach to these attractors can be
examined by considering
$u=2\pi k+\delta u$, with $|\delta u|\ll1$.
At sufficiently late times, where
$H\simeq H_{\rm dS}$, the linearized
axion equation becomes
\begin{equation}
\delta u''
+3\delta u'
+\frac{\mu^2}{\Omega_{\Lambda0}}\delta u
\simeq
-\frac{3\Omega_{{\rm M},*}}
{\tilde f^2\Omega_{\Lambda0}}
\mathcal{A}_{,u}(2\pi k)e^{-3N}.
\label{eq:late_linearized_axion}
\end{equation}
The homogeneous solutions evolve as
$e^{s_\pm N}$, where
$s_\pm=-3/2\pm
\sqrt{9-4\mu^2/\Omega_{\Lambda0}}/2$,
in agreement with the eigenvalues
of Eq.~\eqref{eq:min_desitter_eigenvalues}.
The approach is nonoscillatory for
$\mu^2/\Omega_{\Lambda0}<9/4$ and
oscillatory for $\mu^2/\Omega_{\Lambda0}>9/4$,
with critical damping at equality.

The monopole interaction also induces a
decaying displacement of the axion field.
A particular solution of
Eq.~\eqref{eq:late_linearized_axion} is
\begin{equation}
\delta u_{\rm M}(N)=
-\frac{3\Omega_{{\rm M},*}}
{\tilde f^2\mu^2}
\mathcal{A}_{,u}(2\pi k)e^{-3N},
\label{eq:late_monopole_displacement}
\end{equation}
which, for the quadratic dyon-mass
function, gives
\begin{equation}
\delta u_{\rm M}(N)=
\frac{3\gamma\alpha_D^2
\Omega_{{\rm M},*}}
{2\pi\tilde f^2\mu^2}
(n_e-k)e^{-3N}.
\label{eq:late_monopole_displacement_explicit}
\end{equation}
Hence, the monopole population modifies
the evolution towards the attractor,
but its contribution decreases as
$a^{-3}$ and does not alter the
asymptotic de Sitter state.

\begin{figure}[htbp]
\centering
\includegraphics[width=0.95\columnwidth]
{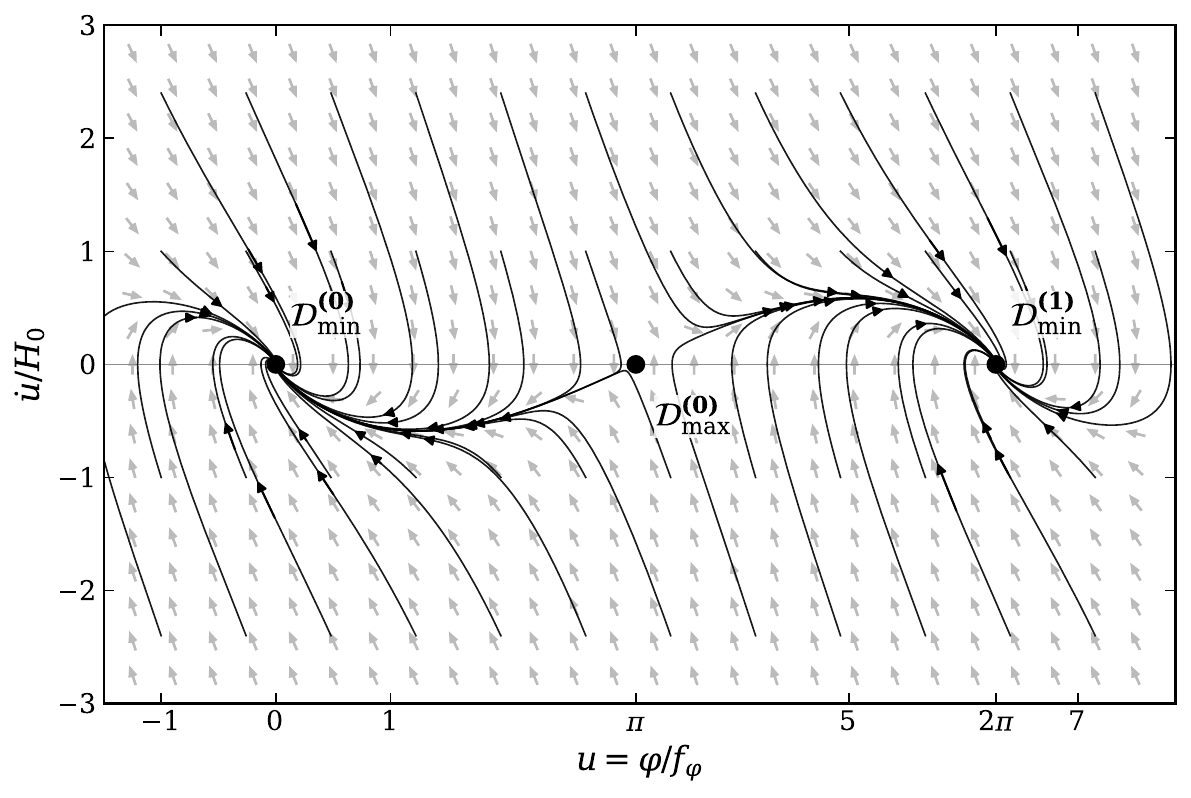}
\caption{{\it{
Phase-space trajectories of the axion-monopole
cosmology in the presence of a positive
cosmological constant, with
$\Omega_{\Lambda,0}=0.38$, where
$\Omega_{\Lambda,0}\equiv
\rho_\Lambda/(3M_{\rm Pl}^2H_0^2)$.
The horizontal and vertical axes represent
$u=\varphi/f_\varphi$ and $\dot u/H_0$,
respectively.
The filled circles denote the de Sitter
critical points.
The points $\mathcal{D}_{\rm min}^{(0)}$
and $\mathcal{D}_{\rm min}^{(1)}$,
at $u=0$ and $u=2\pi$, are stable
late-time attractors, while
$\mathcal{D}_{\rm max}^{(0)}$, at
$u=\pi$, is unstable.
The solid black curves represent
representative cosmological trajectories,
with arrows indicating the direction
of increasing e-fold time $N$.
The light-gray arrows illustrate
instantaneous projected flow directions
for a prescribed choice of the remaining
dynamical variables.
The diagram is a two-dimensional projection
of the full cosmological system.
}}}
\label{fig:phase_positive_lambda}
\end{figure}

The corresponding phase-space behavior
is illustrated in Fig.~\ref{fig:phase_positive_lambda},
where we present trajectories projected
onto the $(u,\dot u/H_0)$ plane for
a positive cosmological constant.
The evolution is characterized by
two stable de Sitter attractors,
$\mathcal{D}_{\rm min}^{(0)}$ and
$\mathcal{D}_{\rm min}^{(1)}$, located
at $u=0$ and $u=2\pi$, respectively.
The intermediate point
$\mathcal{D}_{\rm max}^{(0)}$, at $u=\pi$,
is unstable. Depending on the initial
conditions, trajectories approach one
of the two potential minima, with the
axion velocity gradually vanishing.
The figure thus illustrates the coexistence
of distinct stable de Sitter configurations,
which share the same asymptotic expansion
rate but correspond to different values
of the axion field.

We next consider the case where an explicit cosmological constant is absent, namely
$\Omega_{\Lambda0}=0$.
In this case, the effective dark-energy
sector consists entirely of the axion,
with $\rho_{\rm DE}=\rho_\varphi$ and
$w_{\rm DE}=w_\varphi$.
The de Sitter minima no longer exist at
finite positive $H$, since the axion
potential vanishes there.
The remaining de Sitter solutions,
located at the potential maxima
$u_c=(2k+1)\pi$, satisfy
\begin{equation}
H_{\rm max}^2=
\frac{2m_\varphi^2f_\varphi^2}
{3M_{\rm Pl}^2}.
\label{eq:lambda_zero_desitter}
\end{equation}
However, they possess the positive
eigenvalue
\begin{equation}
s_+=-\frac{3}{2}
+\frac{1}{2}
\sqrt{9+\frac{6}{\tilde f^2}}>0,
\label{eq:lambda_zero_unstable_eigenvalue}
\end{equation}
and therefore cannot represent stable
late-time accelerated attractors.

For trajectories that become trapped
around a potential minimum, the axion
eventually enters an oscillatory regime.
When $H\ll m_\varphi$, the potential
reduces to
$V(\varphi)\simeq
m_\varphi^2(\varphi-\varphi_k)^2/2$,
where $\varphi_k=2\pi kf_\varphi$.
The oscillation amplitude decreases
approximately as $a^{-3/2}$, and
averaging over several oscillations yields
\begin{equation}
\langle p_\varphi\rangle\simeq0,
\qquad
\langle\rho_\varphi\rangle\propto a^{-3},
\qquad
\langle w_{\rm DE}\rangle
=\langle w_\varphi\rangle\simeq0.
\label{eq:late_oscillating_axion}
\end{equation}
Thus, the axion behaves effectively
as pressureless matter, while the
monopole-induced force becomes
progressively negligible relative
to the intrinsic axion dynamics.
For such trajectories, the late-time
evolution is consequently decelerating,
with $
\langle q\rangle\to\frac12$.
This regime corresponds to $H\to0$,
or $y\to\infty$, and therefore lies
outside the finite critical-point
analysis of the previous subsection.

\begin{figure}[htbp]
\centering
\includegraphics[width=0.95\columnwidth]
{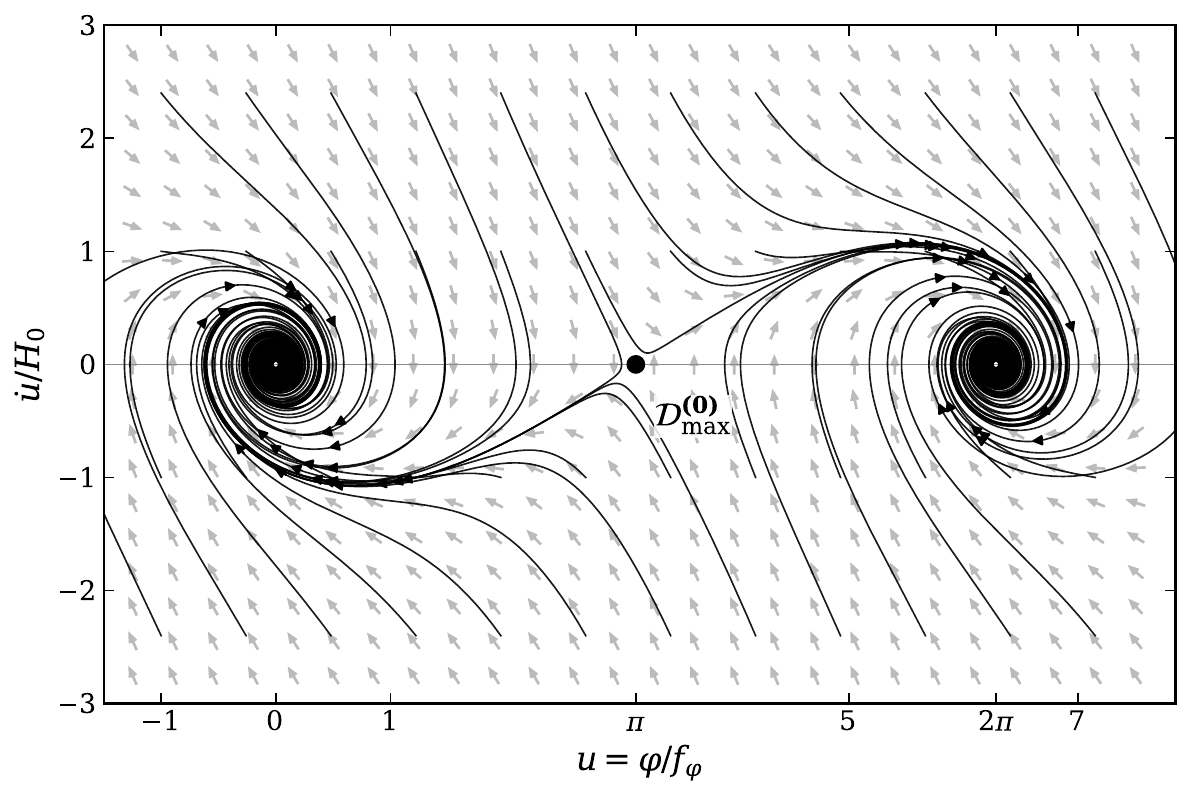}
\caption{{\it{
Phase-space trajectories of the axion-monopole
cosmology in the absence of an explicit
cosmological constant,
$\Omega_{\Lambda,0}=0$.
The horizontal and vertical axes represent
$u=\varphi/f_\varphi$ and $\dot u/H_0$,
respectively.
The filled circle denotes the unstable
de Sitter critical point
$\mathcal{D}_{\rm max}^{(0)}$ at $u=\pi$.
There is no stable finite-Hubble de Sitter
attractor in this case.
The solid black curves illustrate
representative cosmological trajectories,
with arrows indicating increasing
e-fold time $N$.
The inward spirals correspond to damped
oscillations of the axion around the
potential minima at $u=0$ and $u=2\pi$,
which are approached as $H\to0$ and
are not finite-Hubble critical points.
The light-gray arrows indicate
instantaneous projected flow directions
for a prescribed choice of the remaining
dynamical variables.
As in Fig.~\ref{fig:phase_positive_lambda},
the diagram represents a two-dimensional
projection of the full cosmological system.
}}}
\label{fig:phase_zero_lambda}
\end{figure}

This behavior can be seen in
Fig.~\ref{fig:phase_zero_lambda},
which shows the projected phase-space
evolution in the absence of an explicit
cosmological constant.
The de Sitter point
$\mathcal{D}_{\rm max}^{(0)}$ is unstable,
and trajectories departing from its
vicinity may subsequently evolve towards
the neighboring potential minima.
The spiraling trajectories correspond
to damped axion oscillations, whose
amplitudes decrease as the universe
expands. Unlike the case with
$\Omega_{\Lambda0}>0$, these minima do
not represent finite-Hubble de Sitter
critical points. Instead, the corresponding
solutions evolve towards $H\to0$.
The absence of a stable de Sitter
attractor is therefore reflected in
the qualitatively different late-time
behavior of the trajectories.

Nevertheless, the absence of a stable
accelerated attractor does not exclude
a period of axion-driven acceleration.
Using Eqs.~\eqref{eq:epsilon_friedmann}
and \eqref{eq:autonomous_observables},
the acceleration condition reads
\begin{equation}
3x^2+\frac32(v+b)+2r<1.
\label{eq:late_acceleration_criterion}
\end{equation}
Employing the Friedmann constraint,
this can be expressed in terms of
the effective dark-energy density as
\begin{equation}
\Omega_{\rm DE}>
\frac13+2x^2+\frac{r}{3}.
\label{eq:late_acceleration_potential}
\end{equation}
Equivalently, since
$\Omega_{\rm DE}=x^2+\Omega_V+
\Omega_{\Lambda0} y^2$, acceleration
requires
$\Omega_V+\Omega_{\Lambda0} y^2>
1/3+x^2+r/3$.
For $\Omega_{\Lambda0}=0$, this condition
may be satisfied while the axion
evolves slowly along its potential.
Consequently, the axion can dominate
the cosmic energy budget and drive
accelerated expansion even without an
explicit cosmological constant.
The monopole-induced interaction can
modify the field displacement and
velocity acquired during matter domination,
thereby affecting the onset and duration
of this accelerated phase.
However, for trajectories that eventually
settle into the oscillatory regime,
acceleration comes to an end.
Thus, the absence of a stable accelerated
attractor constrains the asymptotic
evolution of the model, without excluding
an axion-dominated accelerating epoch
at finite cosmological times.

Finally, the canonical nature of
the axion field, together with
Eq.~\eqref{eq:late_hubble_monotonicity},
implies
\begin{equation}
\dot H\leq0,\qquad
w_{\rm DE}\geq-1,
\label{eq:no_superacceleration}
\end{equation}
where the second inequality holds
whenever $\rho_{\rm DE}>0$.
Thus, the physical dark-energy sector
does not exhibit phantom behavior.
Nevertheless, the interaction-effective
dark-energy equation-of-state parameter
$w_{\rm DE}^{\rm eff}$, introduced in
Eq.~\eqref{eq:wDE_effective}, can cross
the phantom divide due to the energy
exchange between the axion and monopole
sectors. Such an effective phantom
crossing does not require a phantom
scalar field or imply superacceleration,
since $\dot H\leq0$ remains satisfied.
This distinction is particularly relevant
when interpreting the cosmological
evolution in terms of an effective
dark-energy component
\cite{Bagherian:2026jxz}.
We investigate this behavior numerically
in the next section.

\section{Cosmological evolution and dyon-mass effects}
\label{sec:cosmological_evolution}

Having established the critical points and their stability,
we now investigate the cosmological evolution of the
axion-monopole system. We first examine the Hubble-tracking
behavior during matter domination, deriving approximate
analytical expressions for the axion evolution and the
associated energy transfer. We then solve the background
equations numerically, focusing on the transition to
accelerated expansion and the role of the interaction.
Finally, we examine the sensitivity of the results to
the dyon-mass prescription, comparing the quadratic
approximation with the full BPS expression.

\subsection{Matter-era evolution and Hubble tracking}
\label{subsec:matter_tracking}

An important feature of axion-monopole cosmology
is that the interaction can sustain the evolution of
the axion field during matter domination, even when
its intrinsic potential is dynamically subdominant
\cite{Bagherian:2026jxz}. To investigate this behavior,
we express the axion equation
\eqref{eq:normalized_axion} in terms of the
autonomous variables of Sec. \ref{sec:dynamical_system},  obtaining
\begin{equation}
u''+(3-\varepsilon)u'
+\mu^2y^2\sin u
+\frac{3v}{\tilde f^2}
\frac{\mathcal{A}_{,u}}{\mathcal{A}}=0.
\label{eq:tracking_field_equation}
\end{equation}
The last term describes the monopole-induced force,
whose relative importance depends on the cosmological
epoch. During radiation domination,
$H^2\propto a^{-4}$, whereas the monopole number
density scales as $n_{\rm M}\propto a^{-3}$.
Consequently,
\begin{equation}
\frac{n_{\rm M}}{H^2}\propto a.
\label{eq:radiation_tracking_scaling}
\end{equation}
The interaction is therefore suppressed at early
times. During matter domination, however,
$H^2\propto a^{-3}$, leading to
\begin{equation}
\frac{n_{\rm M}}{H^2}\simeq\mathrm{const.}
\label{eq:matter_tracking_scaling}
\end{equation}
Thus, the monopole-induced force can remain
relevant relative to Hubble friction over
an extended period.

The corresponding effective potential can be
expressed as
\begin{equation}
V_{\rm eff}(u,N)=
m_\varphi^2f_\varphi^2(1-\cos u)
+V_{\rm ind}(u,N),
\label{eq:tracking_effective_potential}
\end{equation}
where
\begin{equation}
V_{\rm ind}(u,N)=
3M_{\rm Pl}^2H_0^2\Omega_{{\rm M},*}e^{-3N}
\left[
\mathcal{A}(u)-\mathcal{A}(2\pi n_e)
\right].
\label{eq:tracking_induced_potential}
\end{equation}
This additional potential describes the force
acting on the axion and should not be included
as an independent contribution to the Friedmann
equation, since the monopole energy is already
contained in $\rho_{\rm M}$.

For the quadratic dyon-mass function, the induced
mass satisfies
\begin{equation}
\frac{m_{\rm ind}^2}{H^2}
=
\frac{3\gamma\alpha_D^2v}
{4\pi^2\tilde f^2\mathcal{A}(u)}.
\label{eq:tracking_mass_ratio}
\end{equation}
Hence, provided that $v$ and $u$ evolve slowly,
the induced mass remains approximately proportional
to the Hubble scale during matter domination.

We can obtain an analytical approximation for
the axion evolution in this regime. Taking
$\varepsilon\simeq3/2$ and $H\gg m_\varphi$,
Eq.~\eqref{eq:tracking_field_equation} reduces to
\begin{equation}
u''+\frac32u'
\simeq
-\frac{3v}{\tilde f^2}
\frac{\mathcal{A}_{,u}}{\mathcal{A}}.
\label{eq:matter_axion_approximation}
\end{equation}
If the right-hand side varies slowly, the
solution for the axion velocity becomes
\begin{equation}
u'(N)\simeq
u'_{\rm tr}
+
\left[u'(N_i)-u'_{\rm tr}\right]
e^{-3(N-N_i)/2},
\label{eq:tracking_velocity_solution}
\end{equation}
where
\begin{equation}
u'_{\rm tr}
=
-\frac{2v}{\tilde f^2}
\frac{\mathcal{A}_{,u}}{\mathcal{A}}
=
\frac{\gamma\alpha_D^2v}
{\pi\tilde f^2\mathcal{A}(u)}
\left(n_e-\frac{u}{2\pi}\right).
\label{eq:tracking_velocity}
\end{equation}
Note that the validity of this approximation requires
the tracking velocity to vary slowly compared
with the Hubble-friction timescale, namely
$
\left|
\frac{d\ln|u'_{\rm tr}|}{dN}
\right|\ll\frac{3}{2}$.
Hence, deviations from the tracking velocity
decay approximately as $a^{-3/2}$, while
the axion continues to evolve under the
monopole-induced force. In particular, the
field moves towards increasing $u$ when
$u<2\pi n_e$, and in the opposite direction
when $u>2\pi n_e$.

Importantly, this tracking regime does not
correspond to an exact critical point of
the autonomous system. Since
$u'_{\rm tr}\neq0$ in general, the field
and the interaction strength continue to
evolve. Rather, tracking represents an
approximate attracting trajectory for the
axion velocity, maintained by Hubble
friction as long as the drift velocity
varies sufficiently slowly.

The interaction also modifies the monopole
energy density. Using Eq.~\eqref{eq:tracking_velocity},
the energy-transfer rate becomes
\begin{equation}
\frac{Q}{H\rho_{\rm M}}
=
\frac{\mathcal{A}_{,u}}{\mathcal{A}}u'
\simeq
-\frac{2v}{\tilde f^2}
\left(
\frac{\mathcal{A}_{,u}}{\mathcal{A}}
\right)^2
\leq0.
\label{eq:tracking_energy_transfer}
\end{equation}
Hence, during monopole-driven tracking,
energy flows from the monopole sector
towards the axion, independently of the
sign of the effective dyon charge.
The direction of energy transfer may
change once the intrinsic axion potential
becomes important.

For illustration, adopting
$\alpha_D=0.044$, $\gamma=1$,
$n_e=30$, $\tilde f=0.461$,
$u_i=3\pi/4$, and a representative
matter-era monopole fraction $v\simeq0.83$,
we obtain
\begin{equation}
\frac{m_{\rm ind}^2}{H^2}
\simeq3.1\times10^{-4},
\qquad
u'_{\rm tr}\simeq3.9\times10^{-2}.
\label{eq:tracking_benchmark}
\end{equation}
We mention that the relatively large value $n_e=30$ is
adopted as an illustrative benchmark
to enhance the effects of the axion-dependent
dyon mass, rather than as a preferred
microscopic charge state.
Thus, despite the small induced mass,
the interaction can sustain a gradual
axion evolution throughout matter
domination. These estimates use the
quadratic mass approximation, whose
validity for highly charged populations
will be examined in  subsection \ref{subsec:bps_dyon_mass}.

As the Universe approaches dark-energy
domination, the monopole density decreases
and the intrinsic axion potential becomes
increasingly relevant. The field consequently
departs from the tracking regime, with its
subsequent evolution depending on the
displacement and velocity accumulated
during matter domination. These effects,
and their implications for the onset of
accelerated expansion, are investigated
numerically in the following subsection.

\subsection{Numerical cosmological evolution}
\label{subsec:numerical_evolution}

We now investigate the cosmological evolution of the
axion-monopole system by numerically integrating
Eqs.~\eqref{eq:normalized_friedmann},
\eqref{eq:normalized_axion}, and
\eqref{eq:normalized_hubble}, imposing the Friedmann
constraint throughout the evolution.

As a representative example, we consider
$\alpha_D=0.044$, $\gamma=1$, $n_e=30$,
$\tilde f=0.461$, $\mu=0.5$, and
$\Omega_{\Lambda0}=0.65$, with initially frozen
axion conditions $u_i=3\pi/4$, $u'_i=0$
at the initial redshift $z_i=10^{14}$. We adopt
$\Omega_{b0}=0.04952$,
$\Omega_{r0}=9.24\times10^{-5}$, and
$\Omega_{{\rm M},*}=0.144668$, ensuring
$E(0)=1$. These values provide an illustrative
realization rather than an observational fit.

\begin{figure}[t]
\centering
\includegraphics[width=0.46\textwidth]
{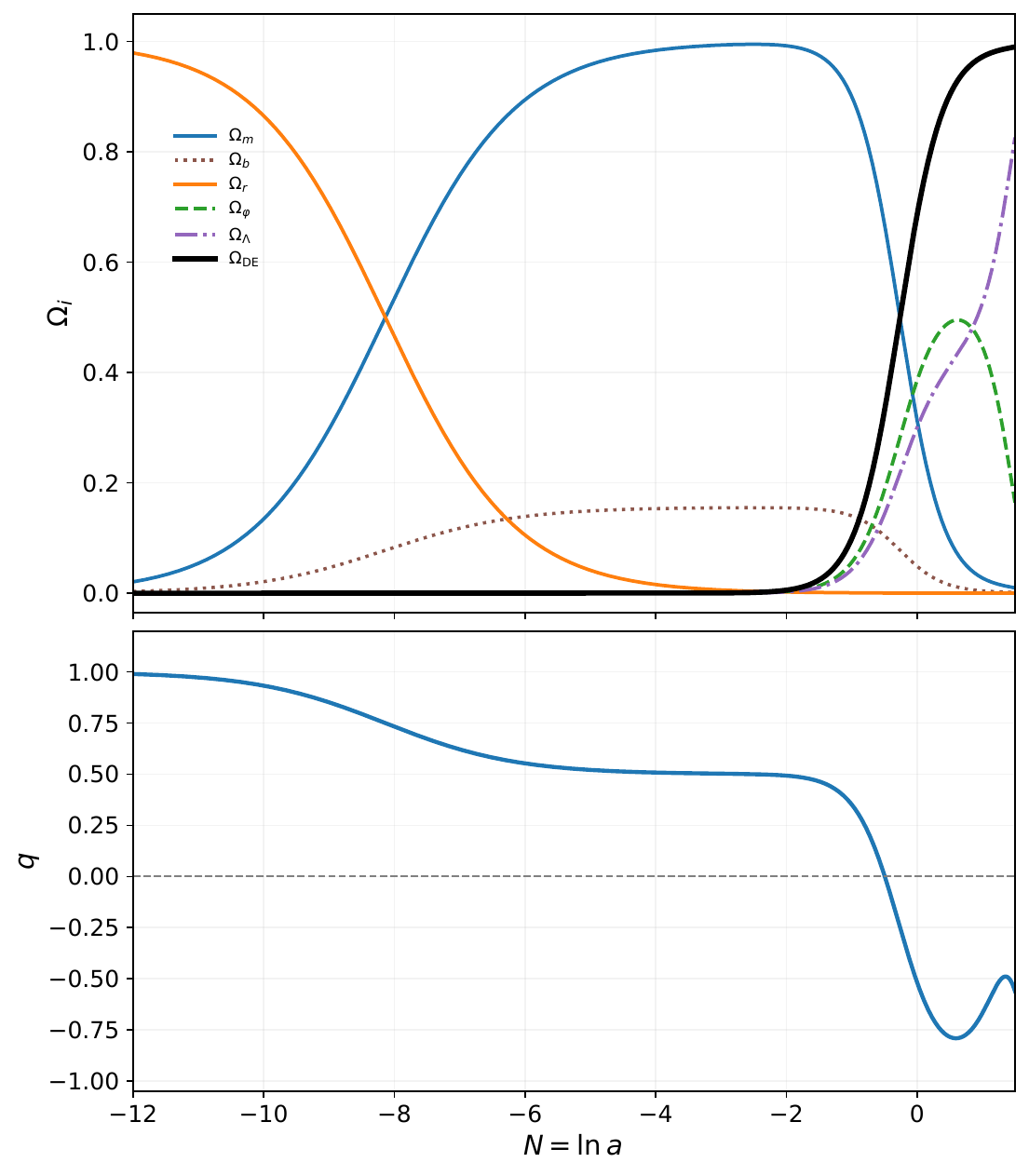}
\caption{\textit{Cosmological evolution of the
axion-monopole model for the representative
parameters given in the text.
Upper panel: evolution of the total matter
($\Omega_m$), baryonic matter ($\Omega_b$),
radiation ($\Omega_r$), axion ($\Omega_\varphi$),
cosmological-constant ($\Omega_\Lambda$),
and total dark-energy ($\Omega_{\rm DE}$)
density parameters, with
$\Omega_m=\Omega_{\rm M}+\Omega_b$ and
$\Omega_{\rm DE}=\Omega_\varphi+\Omega_\Lambda$.
Lower panel: evolution of the deceleration
parameter $q$ as a function of $N=\ln a$.
The transition from decelerated to accelerated
expansion occurs when $q$ crosses zero.}}
\label{fig:axion_background_evolution}
\end{figure}

As shown in Fig.~\ref{fig:axion_background_evolution},
the model reproduces the standard sequence of
radiation, matter, and dark-energy domination.
The monopole component provides the dominant
matter contribution, while the axion and the
cosmological constant jointly determine the
late-time dark-energy density.
For the representative solution,
radiation-matter equality occurs at
$z\simeq3432$, while accelerated expansion
begins at $z_{\rm acc}\simeq0.627$.
The interaction also sustains the matter-era
axion drift discussed in subsection \ref{subsec:matter_tracking} above, with
$u'\simeq0.03883$ at $z=10$, compared with
the analytical estimate $u'_{\rm tr}\simeq0.03899$.

\begin{figure}[t]
\centering
\includegraphics[width=0.48\textwidth]
{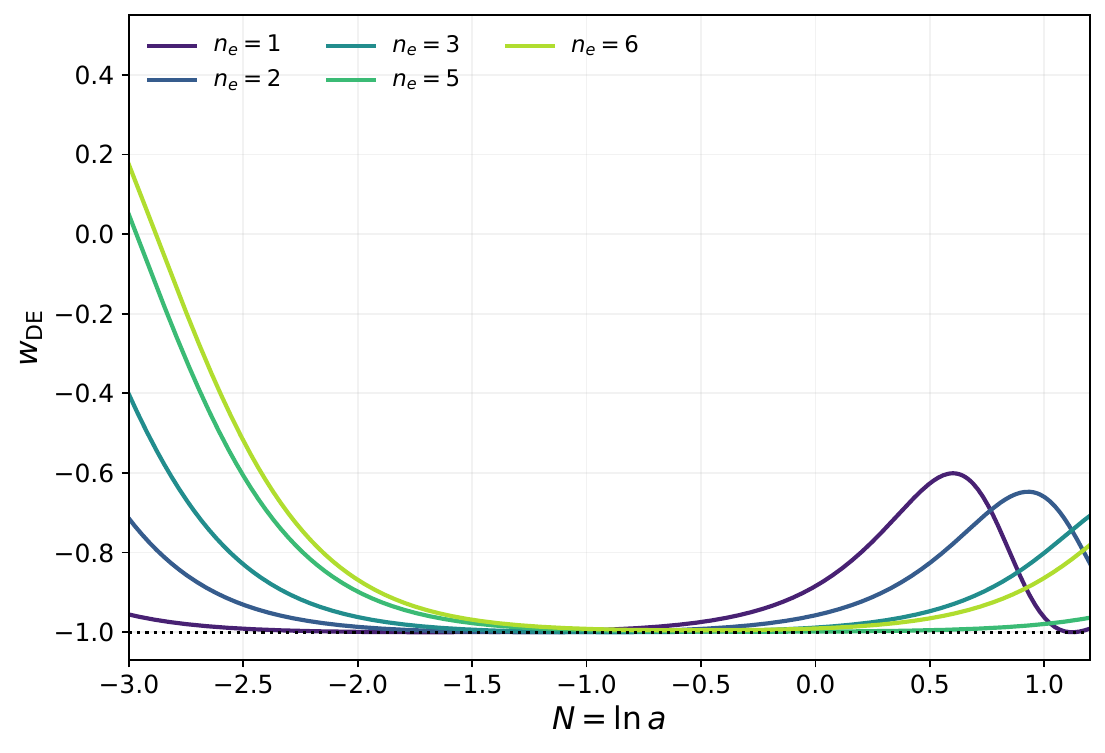}
\caption{\textit{Evolution of the physical
dark-energy equation-of-state parameter
$w_{\rm DE}$ as a function of $N=\ln a$
for five dyon-charge parameters,
$n_e=1,2,3,5,6$.
We adopt $\alpha_D=0.12$, $\tilde f=0.36$,
and $\Omega_{\Lambda,0}=0.3$, adjusting
$\mu$ for each charge state to satisfy
the present-day Friedmann normalization
$E(0)=1$. The other parameters and
initial conditions are kept fixed.
The horizontal dotted line indicates
the phantom divide $w=-1$.
All trajectories remain in the non-phantom
regime, as expected for a canonical axion
combined with a cosmological constant.}}
\label{fig:physical_dark_energy_eos}
\end{figure}

We next examine the dark-energy equation of
state for different dyon populations.
According to Eq.~\eqref{eq:dark_energy_eos},
the physical equation-of-state parameter
satisfies $w_{\rm DE}\geq-1$, since the axion
is canonical and the potential and vacuum
energy are nonnegative. Figure~\ref{fig:physical_dark_energy_eos}
confirms this behavior. Although the
interaction modifies the field evolution,
none of the physical dark-energy trajectories
crosses the phantom divide.

Nevertheless, the interaction-effective
dark-energy equation-of-state parameter
$w_{\rm DE}^{\rm eff}$, defined in
Eq.~\eqref{eq:wDE_effective}, can cross
the phantom divide due to the energy
exchange between the axion and monopole
sectors. In particular, such a crossing
can occur when energy flows from the
monopole sector to the axion.

\begin{figure}[ht!]
\centering
\includegraphics[width=0.48\textwidth]
{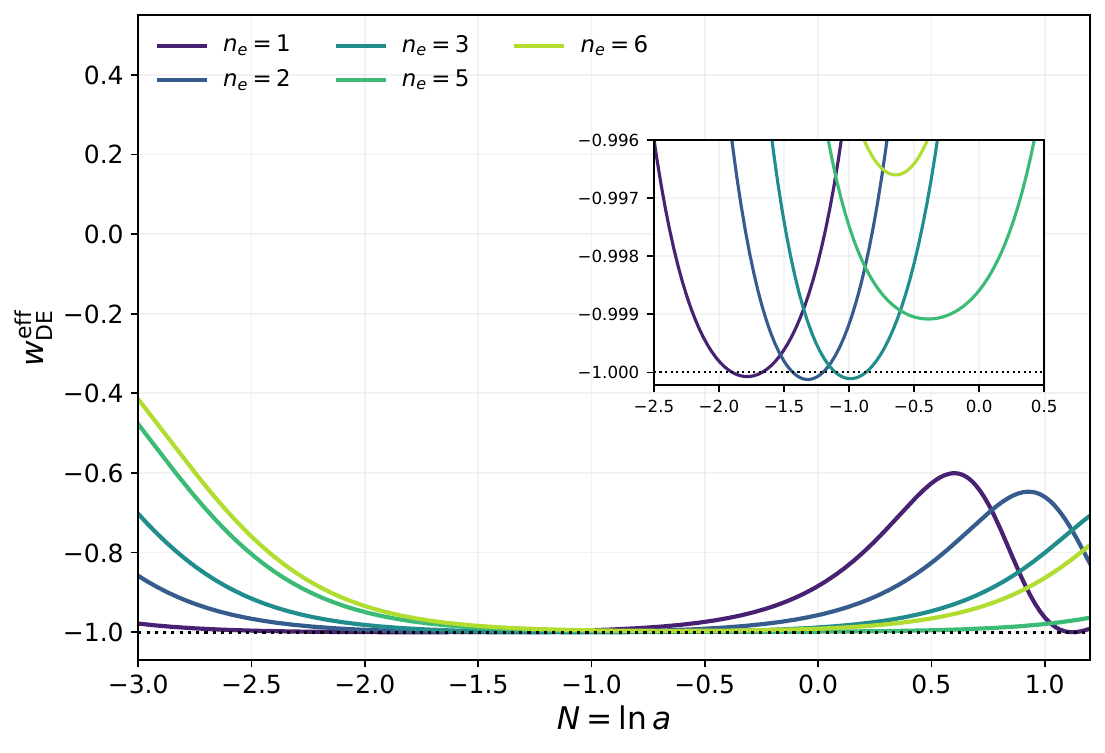}
\caption{\textit{Evolution of the
interaction-effective dark-energy
equation-of-state parameter
$w_{\rm DE}^{\rm eff}$ for the same
parameter choices and initial conditions
as in Fig.~\ref{fig:physical_dark_energy_eos}.
The horizontal dotted line denotes
$w=-1$. The cases $n_e=1,2,3$
exhibit small excursions into the
effective phantom regime, whereas
$n_e=5,6$ remain non-phantom over
the displayed interval.
The inset magnifies the region around
the phantom divide, highlighting the
dependence of the effective crossing
on the dyon-charge parameter.}}
\label{fig:effective_dark_energy_eos}
\end{figure}

In particular, as illustrated in
Fig.~\ref{fig:effective_dark_energy_eos},
the effective equation of state can exhibit
either phantom or non-phantom behavior,
depending on the dyon-charge distribution.
The phantom excursions are small, but
demonstrate that the interaction can
generate an effective crossing without
introducing a phantom scalar field or
violating $\dot H\leq0$.

\subsection{Effects of the full BPS dyon mass}
\label{subsec:bps_dyon_mass}

The preceding analysis employed the quadratic
approximation for the dyon mass. However, for
sufficiently large dyon charges, the expansion
parameter $\alpha_D|n_e-u/(2\pi)|$ becomes
of order unity, and higher-order contributions
can no longer be neglected. We therefore examine
the robustness of the cosmological evolution
using the full Bogomol'nyi-Prasad-Sommerfield
(BPS) mass expression.

In the BPS limit, the mass of a dyon with
electric-charge quantum number $n$ is
\begin{equation}
m_D^{\rm BPS}(n,u)
=m_{\rm M}
\sqrt{1+\alpha_D^2
\left(n-\frac{u}{2\pi}\right)^2}.
\label{eq:bps_dyon_mass}
\end{equation}
For a population occupying a single charge
state $n=n_e$, the mass-modulation function is
\begin{equation}
\mathcal{A}_{\rm BPS}(u)=
\sqrt{1+\alpha_D^2
\left(n_e-\frac{u}{2\pi}\right)^2}.
\label{eq:bps_mass_function}
\end{equation}
Expanding for $\alpha_D|n_e-u/(2\pi)|\ll1$
reproduces Eq.~\eqref{eq:mass_function_single}
with $\gamma=1$.

The logarithmic coupling governing the energy
exchange between the axion and monopole sectors
becomes
\begin{equation}
\mathcal{B}_{\rm BPS}(u)=
-\frac{\alpha_D^2}{2\pi\tilde f}
\frac{n_e-u/(2\pi)}
{1+\alpha_D^2
\left(n_e-u/(2\pi)\right)^2}.
\label{eq:bps_coupling}
\end{equation}
Hence, comparing with the quadratic approximation,
we obtain
\begin{equation}
\frac{\mathcal{B}_{\rm BPS}}
{\mathcal{B}_{\rm quad}}
=
\frac{1+\eta/2}{1+\eta},
\qquad
\eta\equiv\alpha_D^2
\left(n_e-\frac{u}{2\pi}\right)^2.
\label{eq:bps_coupling_ratio}
\end{equation}
Thus, for the same field value and charge state,
the full BPS prescription predicts a weaker
logarithmic coupling. For
$\alpha_D=0.044$, $n_e=30$, and $u_i=3\pi/4$,
we find $\eta_i\simeq1.699$ and
$\mathcal{B}_{\rm BPS}/
\mathcal{B}_{\rm quad}\simeq0.685$,
indicating a significant overestimation of
the interaction strength by the quadratic
approximation.

The curvature of the monopole-induced
potential is also modified, namely
\begin{equation}
\frac{m_{\rm ind,BPS}^2}{H^2}
=
\frac{3v\alpha_D^2}
{4\pi^2\tilde f^2}
\left[
1+\alpha_D^2
\left(n_e-\frac{u}{2\pi}\right)^2
\right]^{-2}.
\label{eq:bps_induced_mass_ratio}
\end{equation}
This contribution remains positive but
is suppressed at large charge offsets.
Therefore, the interaction retains its
restoring character, although the
matter-era axion drift is reduced.

To quantify these effects, we compare the
two mass prescriptions at fixed parameters,
adopting $\alpha_D=0.044$, $n_e=30$,
$\tilde f=0.461$, $\mu=0.5$, and
$u_i=3\pi/4$, with $u'_i=0$ at
$z_i=10^{14}$. We impose the same present
monopole density,
$\Omega_{{\rm M},0}=0.267208$, and
$E(0)=1$ in both cases.
The corresponding normalization gives
$\Omega_{{\rm M},*}=0.144668$ and
$\Omega_{\Lambda0}=0.650000$ for the
quadratic mass, compared with
$\Omega_{{\rm M},*}=0.162744$ and
$\Omega_{\Lambda0}=0.650871$ for the
full BPS expression. The slight adjustment
of $\Omega_{\Lambda0}$ ensures the
present Friedmann constraint.

For this fixed-charge comparison, the
quadratic approximation gives
$u'\simeq0.03883$ at $z=10$,
whereas the full BPS expression yields
$u'\simeq0.02659$, approximately
$31.5\%$ smaller. Both values remain
close to their respective matter-era
tracking estimates, $0.03899$ and
$0.02673$, confirming that tracking
persists beyond the quadratic approximation.
The axion reverses direction at
$z_{\rm turn}\simeq0.254$ in the
quadratic case, compared with
$z_{\rm turn}\simeq0.512$ for the BPS
mass. By contrast, the background
expansion remains nearly unchanged,
with $z_{\rm acc}\simeq0.62734$ and
$0.62740$, respectively, and
$q_0\simeq-0.52470$ and $-0.52465$.
This near coincidence reflects the
dominance of the cosmological constant
and the matching of the present
matter density.

\begin{figure}[t]
\centering
\includegraphics[width=0.48\textwidth]
{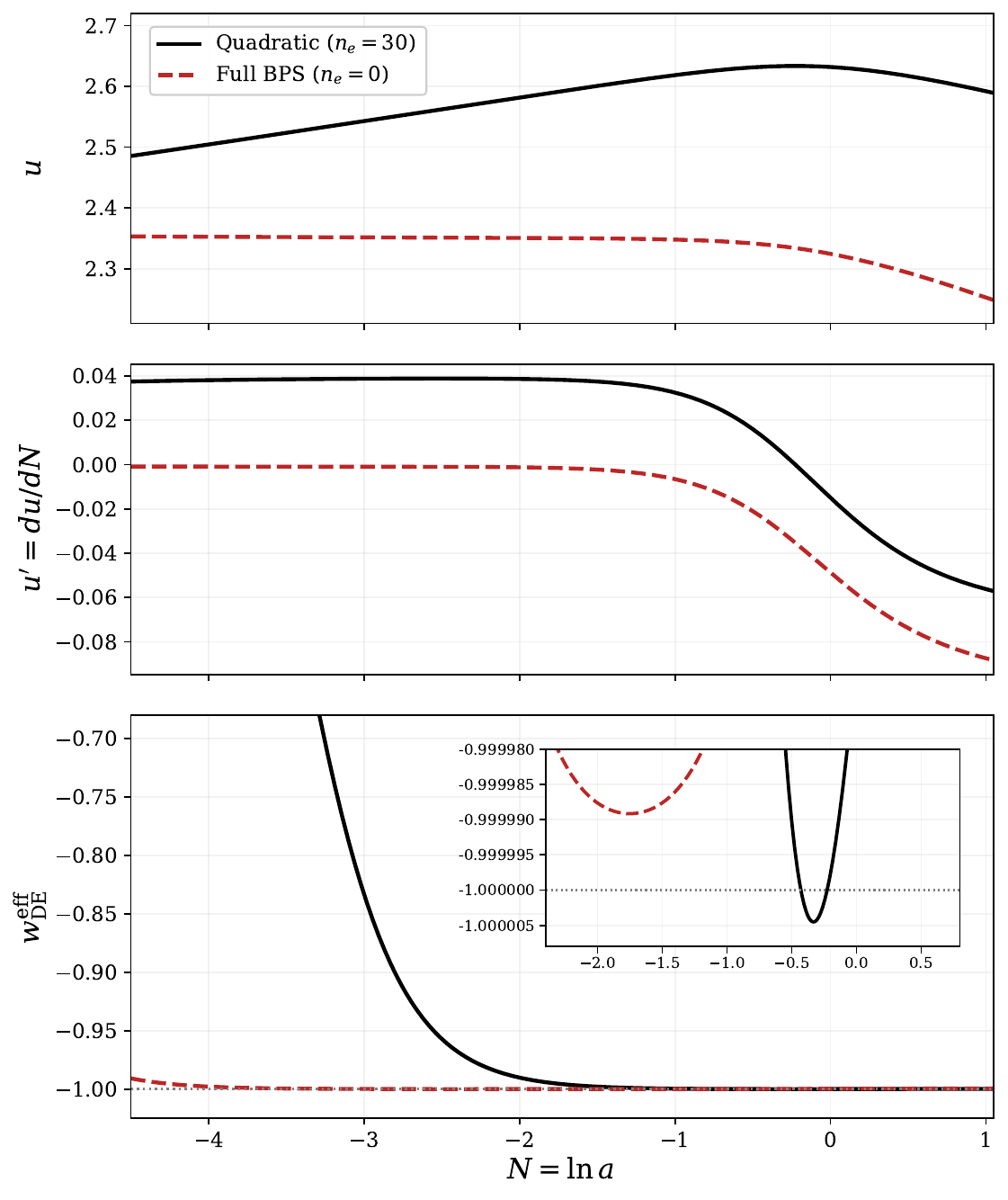}
\caption{\textit{Comparison of the cosmological
evolution for the quadratic dyon-mass
approximation with $n_e=30$ (solid black
curves) and the full BPS mass with $n_e=0$
(dashed red curves).
The upper and middle panels show the
dimensionless axion field $u$ and its
derivative $u'=du/dN$, respectively,
with $N=\ln a$.
The lower panel presents the
interaction-effective dark-energy
equation-of-state parameter
$w_{\rm DE}^{\rm eff}$.
The inset magnifies the region around
the phantom divide $w=-1$, revealing
effective phantom behavior for the
quadratic case and non-phantom behavior
for the BPS case.
Both solutions have the same initial
axion configuration but different
dyon charge states.}}
\label{fig:bps_comparison}
\end{figure}

We further illustrate the effects of
the mass prescription and dyon charge
in Fig.~\ref{fig:bps_comparison},
comparing the quadratic mass with
$n_e=30$ and the full BPS mass with
$n_e=0$. Both cases share the same
initial axion configuration.
Since the charge state also differs,
this comparison is illustrative and
should be distinguished from the
fixed-charge analysis above.

As shown in Fig.~\ref{fig:bps_comparison},
the upper and middle panels illustrate
the differences in the accumulated axion
displacement and velocity, with the latter
making the matter-era drift and subsequent
field reversal particularly evident.
The lower panel displays
$w_{\rm DE}^{\rm eff}$, defined in
Eq.~\eqref{eq:wDE_effective}.
The quadratic case with $n_e=30$
exhibits a small effective phantom
excursion, whereas the BPS case with
$n_e=0$ remains non-phantom.
The inset highlights these small
departures from $-1$.

As discussed in  subsection \ref{subsec:numerical_evolution}, this phantom
behavior is effective,
since the canonical axion satisfies
$w_{\rm DE}\geq-1$. Moreover, the
contrasting behavior in
Fig.~\ref{fig:bps_comparison}
cannot be attributed solely to the
mass prescription, since the charge
parameter also changes.
Indeed, at fixed $n_e=30$, both
prescriptions exhibit a small effective
phantom excursion. Thus, the full BPS
mass modifies the interaction and field
evolution without eliminating this
phenomenon.

Finally, the full BPS prescription
generalizes to a distribution of dyon
charge states as
\begin{equation}
\mathcal{A}_{\rm BPS}(u)=
\sum_n f_n
\sqrt{1+\alpha_D^2
\left(n-\frac{u}{2\pi}\right)^2},
\label{eq:bps_charge_distribution}
\end{equation}
where $\sum_n f_n=1$.
Unlike the quadratic approximation,
this expression depends on the full
charge distribution rather than only
its mean and variance.
The minimum of the averaged mass
function satisfies
\begin{equation}
\sum_n f_n
\frac{n-u_*/(2\pi)}
{\sqrt{1+\alpha_D^2
(n-u_*/(2\pi))^2}}=0,
\label{eq:bps_distribution_minimum}
\end{equation}
and need not coincide with
$u_*=2\pi n_e$. Consequently,
the monopole-dominated matter critical
point may shift relative to the
quadratic treatment, while the de Sitter
critical points remain unchanged
because their monopole density vanishes.
Note that for a charge distribution symmetric about $n_e$,
the minimum remains at $u_*=2\pi n_e$,
whereas asymmetric distributions can shift
its position. Nevertheless, the coupling
vanishes at the shifted minimum,
$\mathcal{B}_{\rm BPS}(u_*)=0$.
Thus, the corresponding matter-dominated
critical point still has vanishing energy
transfer, as in the quadratic treatment.

In summary, the full BPS prescription
preserves the matter-era tracking
mechanism and the late-time de Sitter
attractors, while modifying the
interaction strength, axion displacement,
and field-reversal epoch.
The effective dark-energy behavior
also depends on the dyon charge state,
highlighting the role of microscopic
dyon properties in the cosmological
evolution of the interacting dark sector.

\section{Conclusions}
\label{Conclusions}

The nature of dark matter and dark energy, and the
possibility of interactions between them, remain open
questions in cosmology. Most interacting dark-energy
models introduce the coupling phenomenologically,
without specifying the underlying particle-physics
mechanism. The axion-monopole scenario offers a
different approach, since the Witten effect renders
dyon masses dependent on the axion field and thereby
generates an exchange of energy between the dark
sectors. This motivates a systematic investigation
of the resulting cosmological dynamics, in particular
whether the interaction can sustain stable accelerated
expansion and how it affects the evolution of dark
energy.

In this work, we investigated the cosmological
dynamics of an ultralight canonical axion interacting
with magnetic-monopole dark matter through
field-dependent dyon masses. Considering both the
presence and absence of a cosmological constant,
we formulated the background equations as a closed
autonomous system, identified its finite critical
points, and analyzed their stability and the
conditions for accelerated expansion. We further
studied the matter-era tracking regime analytically
and numerically, examined the physical and
interaction-effective dark-energy equations of state,
and extended the analysis from the quadratic
dyon-mass approximation to the full
Bogomol'nyi-Prasad-Sommerfield (BPS) expression.

We found that the only stable de Sitter critical
points at finite positive Hubble parameter
correspond to minima of the axion potential and
require a positive cosmological constant. At these
attractors the axion kinetic and potential energies
vanish asymptotically, and the expansion is sustained
entirely by the constant vacuum-energy contribution.
The de Sitter configurations associated with
potential maxima, although supported in part by
the axion potential and present even without a
cosmological constant, are unstable. Thus, within
the finite-Hubble phase-space analysis, the
axion-monopole interaction does not generate a
new stable accelerated attractor in the absence
of a cosmological constant. Nevertheless, transient
axion-driven acceleration remains possible, and
its onset and duration can be affected by the
interaction. This shows that the coupling can alter
the cosmological evolution without changing the
nature of its stable de Sitter asymptotic states.

We also clarified the dynamical origin of the
matter-era tracking behavior. The monopole-induced
force sustains a slowly evolving axion, whose
velocity approaches an approximately attracting
slow-drift solution rather than an exact critical
point of the autonomous system. During this regime,
energy flows from the monopole sector to the axion,
while the subsequent field evolution reflects the
competition between the interaction and the intrinsic
axion potential.
Moreover, the interaction-effective dark-energy
equation of state can cross the phantom divide,
even though the physical equation of state
remains non-phantom and the background
satisfies $\dot H\leq 0$. Such behavior therefore
arises from energy exchange, without requiring a
phantom degree of freedom. Using the full BPS dyon
mass, we found that the tracking regime and stable
de Sitter attractors persist, although the
interaction strength, axion displacement, and
field-reversal epoch can differ appreciably from
the quadratic approximation. These differences
highlight the role of microscopic dyon properties
in determining the evolution of the interacting
dark sector.

A natural next step is to extend the analysis to
cosmological perturbations, investigating the
consequences for structure growth, the cosmic
microwave background, and the constraints imposed
by observational data. It would also be of interest
to relax the assumption of fixed dyon charge fractions
and study possible charge-state transitions and
their backreaction on the axion field. More general
axion potentials and monopole populations, together
with a separate analysis of the asymptotic regime
$H\to0$, may reveal additional cosmological behavior
beyond the finite critical points considered here.
The present results thus provide a dynamical basis
for assessing axion-monopole interactions as a
microscopic dark-energy mechanism, identifying both
their characteristic cosmological effects and the
limits of their ability to sustain late-time
acceleration.

 \begin{acknowledgments}  The author acknowledges the contributions of the LISA
CosWG  and  COST
Actions
 CA21106 ``COSMIC WISPers
in the Dark Universe: Theory, astrophysics and experiments'',  CA21136
``Addressing observational tensions in cosmology with
  systematics and fundamental physics (CosmoVerse)'',    CA23130
``Bridging high and low energies in
search of quantum gravity (BridgeQG)'', and CA24101 ``Testing Fundamental
Physics with Seismology''.
 \end{acknowledgments}
 
 %%%%%%%%%%%%%%%%%%%%%%%%%%%%%%%%%%%%%%%%%%%%%%%%%%%%%%%%%%%%%
 
\end{document}